\documentclass[aps,prb,twocolumn,superscriptaddress,showpacs]{revtex4-1}

\usepackage{amsmath}
\usepackage{graphicx}
\usepackage{color}

\newcommand\br{\mathbf{r}}
\newcommand\bk{\mathbf{k}}

\begin{document}

\title{Wigner-function formalism applied to semiconductor quantum devices: \\
Need for nonlocal scattering models}

\author{Rita Claudia Iotti}
\affiliation{
Department of Applied Science and Technology, Politecnico di Torino \\
Corso Duca degli Abruzzi 24, 10129 Torino, Italy
}

\author{Fabrizio Dolcini}
\affiliation{
Department of Applied Science and Technology, Politecnico di Torino \\
Corso Duca degli Abruzzi 24, 10129 Torino, Italy
}
\affiliation{CNR-SPIN, Monte S. Angelo - via Cinthia, I-80126 Napoli, Italy}

\author{Fausto Rossi}
\affiliation{
Department of Applied Science and Technology, Politecnico di Torino \\
Corso Duca degli Abruzzi 24, 10129 Torino, Italy
}

\date{\today}

\begin{abstract}
In designing and optimizing new-generation nanomaterials and related quantum devices, dissipation versus decoherence phenomena are often accounted for via local scattering models, such as relaxation-time and Boltzmann-like schemes. Here we show that the use of such local scattering approaches within the Wigner-function formalism may lead to unphysical results, namely anomalous suppression of intersubband relaxation, incorrect thermalization dynamics, and violation of probability-density positivity. Furthermore, we propose a quantum-mechanical generalization of relaxation-time and Boltzmann-like models, resulting in nonlocal scattering superoperators that enable one to overcome such limitations.
\end{abstract}



\maketitle

\section{Introduction}\label{s-I}

The application of the semiclassical Boltzmann transport theory\cite{b-Jacoboni89} is controversial in various situations of state-of-the-art micro- and nanoelectronics. Indeed, space and/or time scale reduction, enabled by present-day technological advances, pushes new-generation semiconductor devices toward quantum regimes,\cite{b-Bastard88,Frensley90a,Lent93a,DiCarlo94a,Savasta96a,Fischetti99a,Datta00a,Rossi02b,Axt04a,Jacoboni04a,Pecchia04a,Iotti05b} thereby entailing important consequences. 
The first one is the development of quantum approaches, which can be qualitatively grouped in two main classes: double-time approaches based on the nonequilibrium Green's function technique,\cite{b-Datta05,b-Haug07,b-Jacoboni10} and single-time approaches based on the density-matrix theory,\cite{b-Haug04,b-Rossi11} including phase-space treatments within the Wigner-function formalism.\cite{b-Buot09} 
The second consequence is a significant increase in computational effort and resources; indeed, a microscopic treatment of various scattering mechanisms via proper Monte Carlo simulations\cite{b-Jacoboni89} is often computationally too demanding already within the conventional Boltzmann transport theory, and for quantum-transport simulation strategies the situation is even worse. 
Simplified schemes are therefore often adopted to handily design and optimize new-generation nanomaterials and related devices.
Under some conditions, however, such simplified schemes may lead to unrealistic results. For instance, within conventional Wigner-function simulations, the exchange of particles between an open quantum system and its external charge reservoirs is often modeled in terms of semiclassical spatial boundary-condition schemes; This may give rise to non-unique and/or to unphysical solutions, as has been pointed out in Ref.~[\onlinecite{Rosati13a}].

Another important issue in the Wigner-function-based modelling is represented by the simulation of dissipation and decoherence phenomena, which is often performed via simplified local scattering models, namely relaxation-time approximation (RTA) and Boltzmann-like treatments. This paper is devoted to this fundamental aspect. Our goal is twofold: 
(i) we shall show that a naive incorporation of local scattering models within conventional Wigner-function simulation schemes may lead to unphysical results, like anomalous suppression of intersubband relaxation, incorrect thermalization dynamics, and violation of probability-density positivity; 
(ii) we shall propose quantum-mechanical generalizations of conventional relaxation-time as well as Boltzmann-like models, resulting in nonlocal and positivity-preserving scattering superoperators able to overcome the unphysical behaviors just mentioned. In particular, starting from the density-matrix treatment originally proposed in Ref.~[\onlinecite{Taj09b}] and recently extended in Ref.~[\onlinecite{Rosati14e}], we shall derive a corresponding nonlinear Wigner-function scattering superoperator able to describe dissipation and decoherence also in quantum devices operating at high carrier densities.

The paper is organized as follows: In Sec.~\ref{s-WFDM} we briefly recall the main aspects of the Wigner-function formalism, pointing out where locality assumptions arise, and comparing the latter to its density-matrix counterpart.  
In Sec.~\ref{s-LSM}  we  show the intrinsic limitations of relaxation-time and Boltzmann-like  scattering models. Then, in Sec.~\ref{s-NLA} we propose corresponding nonlocal generalizations that allow one to overcome these problems.  Finally, in Sec.~\ref{s-SC} we  summarize our results and draw the conclusions.

\section{Wigner-function versus density-matrix formalism}\label{s-WFDM}
The Wigner-function formalism\cite{Frensley90a,Jacoboni04a} has been adopted in various contexts to study quantum-transport phenomena in semiconductor nanomaterials and nanodevices.\cite{Frensley86a,Kluksdahl89a,Buot90a,Jensen90a,Miller91a,McLennan91a,Tso91a,Gullapalli94a,Fernando95a,ElSayed98a,Kim01a,Pascoli98a,Nedjalkov04a,Nedjalkov06a,Weetman07a,Querlioz08a,Morandi09a,Wojcik09a,Barraud09a,Yoder10a,Alvaro10a,Savio11a,Trovato11a,Sellier14a,Sellier14b,Jonasson15a,Hamerly15a,Cabrera15a,Kim15a} In what follows, we briefly summarize its main features with the purpose of better illustrating our results reported in the next sections.
The Wigner function $f(\br,\bk)$, despite being defined on the classical phase-space $(\br,\bk)$, fully characterizes the quantum and statistical state of the electron subsystem, since it is in one-to-one correspondence with the single-particle density matrix $\hat{\rho}$ via the Weyl-Wigner transform $f(\br,\bk) = {\rm tr}[\hat{W}(\br,\bk)\hat{\rho}]$. It is thus an extremely useful tool offering a twofold advantage. On the one hand, as a function defined on the single-particle phase-space coordinates, it is more intuitive, visualizable and handable than quantities characterizing other quantum formalisms, like the nonequilibrium Green's functions\cite{b-Datta05,b-Haug07,b-Jacoboni10}, whose double-time nature makes their physical interpretation less straightforward and their computation more demanding.
On the other hand, the Wigner function can be regarded as the quantum-mechanical generalization of the conventional distribution function, thus providing a straightforward way to compare with any semiclassical or Boltzmann approach.

The Wigner function fulfills the Wigner transport equation, which may be schematically written as 
\begin{equation}\label{WTE}
\frac{\partial f(\br,\bk)}{\partial t}
=
\left.\frac{\partial f(\br,\bk)}{\partial t}\right|_{\rm d}
+
\left.\frac{\partial f(\br,\bk)}{\partial t}\right|_{\rm s}\ ,
\end{equation}
and is formally reminiscent of the Boltzmann equation for the distribution function. Such basic link has also stimulated the development of so-called Wigner Monte Carlo schemes,\cite{Pascoli98a,Nedjalkov04a} namely simulation techniques based on a Monte Carlo solution of the Wigner transport equation.
The first term on the right-hand side of Eq.~(\ref{WTE}) is  the quantum-mechanical generalization  of the deterministic (diffusion-plus-drift) term in the semiclassical theory, and can be conveniently expressed in terms of the well-known Moyal brackets,\cite{Moyal49a} whose explicit form depends on the electron band dispersion and on the electromagnetic gauge.
Within the customary approximation of a parabolic dispersion characterized by an effective-mass $m^*$, and in the presence of static electric fields only (i.e., no magnetic field) described by a purely scalar potential $V(\br)$, the deterministic (d) contribution to Eq.(\ref{WTE}) is given by
\begin{equation}\label{WTEd}
\left.\frac{\partial f(\br,\bk)}{\partial t}\right|_{\rm d}
=
- \frac{\hbar \bk}{m^*} \cdot {\bf \nabla}_{\br} f(\br,\bk)
- \int d\bk'\,\mathcal{V}(\br,\bk-\bk') f(\br,\bk')\ ,
\end{equation}
where 
\begin{equation}\label{VWW}
\mathcal{V}(\br,\bk'') \!=\! \frac{\imath}{\hbar} \int d\br'
\frac{e^{-\imath \bk'' \cdot \br'}}{2\pi}
\left[\!V\left(\br+\frac{\br'}{2}\right)\!-\!V\left(\br-\frac{\br'}{2}\right)\!\right] 
\end{equation}
is the nonlocal Weyl-Wigner superoperator corresponding to the scalar-potential profile\cite{b-Rossi11}.

The second term in Eq.~(\ref{WTE}) describes energy dissipation and decoherence phenomena induced by various scattering mechanisms. Within a fully quantum-mechanical treatment, such scattering term is strictly nonlocal, as described in detail in Ref.~[\onlinecite{Rosati14b}], and can be written in the form
\begin{equation}\label{NLSS}
\left.\frac{\partial f(\br,\bk)}{\partial t}\right|_{\rm s}
=
S\left[f(\br',\bk')\right](\br,\bk)\ ,
\end{equation}
where, in general, $S$ is a nonlinear scattering superoperator describing a nonlocal action both in $\br$ and $\bk$, i.e., the scattering contribution to the generic phase-space point $(\br,\bk)$ depends on the value of the Wigner function $f$ in any other phase-space point $(\br',\bk')$.

Due to the difficulty in dealing with its fully nonlocal character, it is common practice in many quantum-simulation approaches to replace the scattering superoperator in (\ref{NLSS}) with  a local superoperator.
The simplest choice\cite{Kluksdahl89a,Jensen90a,Wojcik09a,Jonasson15a} is the adoption of a RTA model that rewords the semiclassical case, i.e., 
\begin{equation}\label{RTA}
\left.\frac{\partial f(\br,\bk)}{\partial t}\right|_{\rm s}
=
-\Gamma(\br,\bk)\,\left(f(\br,\bk)-f^\circ(\br,\bk)\right)\ ,
\end{equation}
where  the relaxation of a state $(\br,\bk)$ toward the equilibrium Wigner function $f^\circ(\br,\bk)$ is described in terms of a space- and momentum-dependent relaxation rate $\Gamma(\br,\bk)$ that purely depends on {\em that} state and encodes all relevant scattering processes characterizing the operational conditions of the device.
The space- and energy-dependence of $\Gamma(\br,\bk)$ may be extracted from fully microscopic Monte Carlo simulations,\cite{b-Jacoboni89} or modelled via simplified Fermi's Golden rule treatments.

Another simplified (i.e. local) version of the scattering superoperator in Eq.~(\ref{NLSS}) is inspired by the formal analogy between the Wigner transport equation (\ref{WTE}) and the usual Boltzmann transport theory, and consists in replacing $S$ with a conventional (i.e., semiclassical) Boltzmann collision term\cite{b-Jacoboni89,b-Rossi11,b-Jacoboni10}  
\begin{equation}\label{BTE1}
\left.\frac{\partial f(\br,\bk)}{\partial t}\right|_{\rm s}
\!=\!
\int d\bk'
\left[
P(\br;\bk,\bk') f(\br,\bk')
\!-\!
P(\br;\bk',\bk) f(\br,\bk)
\right]
\end{equation}
where
\begin{equation}\label{BTE2}
P(\br;\bk,\bk') 
=
\left(1-f(\br,\bk)\right)
P_0(\br;\bk,\bk')
\end{equation}
denotes the low-density scattering rate $P_0$ in $\br$ for the generic transition $\bk' \to \bk$, weighted by the usual Pauli-blocking factor, and simply reduces to $P_0(\br;\bk,\bk')$ in the low-density limit ($f(\br,\bk) \to 0$).

Importantly, all such simplified approaches are {\em local in space}; more precisely, the following approximation is made
\begin{equation}\label{LSS}
\left.\frac{\partial f(\br,\bk)}{\partial t}\right|_{\rm s}
 \simeq \,
\bar{S}\left[f(\br,\bk')\right](\br,\bk)\ ,
\end{equation}
i.e., the scattering contribution to the generic phase-space point $\br,\bk$ is assumed to depend on the value of the Wigner function $f$ in $\br$ only.\\

In order to better illustrate this aspect one can focus on the low-density limit, where the original nonlinear scattering superoperator $S$ in (\ref{NLSS}) acquires the form
\begin{equation}\label{LDSM}
\left.\frac{\partial f(\br,\bk)}{\partial t}\right|_{\rm s}
=
\int d\br' d\bk' A(\br,\bk;\br',\bk') f(\br',\bk') + B(\br,\bk)\ .
\end{equation}
Under the approximation scheme of locality in space, 
$A(\br,\bk;\br',\bk') \simeq  \delta(\br- \br') \bar{A}(\br;\bk,\bk')$. Equation~(\ref{LDSM}) therefore reduces~to
\begin{equation}\label{LDSMbis}
\left.\frac{\partial f(\br,\bk)}{\partial t}\right|_{\rm s}
=
\int d\bk' \bar{A}(\br;\bk,\bk') f(\br,\bk') + B(\br,\bk)
\end{equation}
and the semiclassical Boltzmann collision term (\ref{BTE1}) is then recovered by identifying $B(\br,\bk)=0$ and 
\begin{equation}\label{barA-BTE}
\bar{A}(\br;\bk,\bk')=P(\br;\bk,\bk')\,-\,  \delta(\bk -\bk')\int d \bk^{\prime \prime} P(\br;\bk^{\prime \prime},\bk) \, .
\end{equation}
The first term in Eq.~(\ref{barA-BTE}) is local in $\br$ and represents the in-scattering part, while the second one is local both in $\br$ and $\bk$ and corresponds to the out-scattering part. Equation~(\ref{LDSMbis}) may be further simplified by neglecting nonlocal contributions in $\bk'$, i.e. by assuming $\bar{A}(\br;\bk,\bk') \simeq \delta(\bk - \bk') \bar{\bar{A}}(\br,\bk) $. One then obtains
\begin{equation}\label{LDSMter}
\left.\frac{\partial f(\br,\bk)}{\partial t}\right|_{\rm s}
=
\bar{\bar{A}}(\br,\bk) f(\br,\bk) + B(\br,\bk)\quad.
\end{equation}
and the RTA model in (\ref{RTA}) is then recovered upon identifying $\bar{\bar{A}}(\br,\bk)=-\Gamma(\br, \bk)$ and $B(\br,\bk)=\Gamma(\br,\bk) f^\circ(\br,\bk)$.\\

Although approximations are often unavoidable, the degree of accuracy of a given simplified form for the scattering superoperator is intimately related to the nanosystem as well as to the specific phenomenon under investigation. Two basic requirements are, however, always mandatory:
\begin{itemize}
\item[(i)\ ] 
the scattering superoperator should preserve the positive-definite character of our quantum-mechanical state at any time;
\item[(ii)\ ] 
in the absence of external electro-optical excitations the steady-state solution of the transport equation (\ref{WTE}) should coincide with the thermal-equilibrium state.
\end{itemize}
\par\noindent
Checking the fulfillment of these basic requirements is in general a highly non-trivial task; to this aim it is worth exploiting the close relation between the Wigner function and the corresponding single-particle density matrix.\cite{Rossi02b,Iotti05b}
More specifically, adopting the very same notation employed in Ref.~[\onlinecite{Rosati13a}], and by compactly labelling with~$\alpha$ the set of relevant quantum numbers for the single-particle electronic states of  a semiconductor nanostructure, the density matrix $\rho_{\alpha_1\alpha_2}$\footnote{
The diagonal terms of the density matrix describe the population of the generic single-particle state~$\alpha$ while the off-diagonal terms describe the quantum-mechanical phase coherence (or polarization) between states $\alpha_1$ and $\alpha_2$.}
can be written in terms of the Wigner function $f(\br,\bk)$ as\cite{b-Rossi11}
\begin{equation}\label{WWTinv}
\rho_{\alpha_1\alpha_2} = \int \frac{d\br\,d\bk}{(2\pi)^3} W^{ }_{\alpha_1\alpha_2}(\br,\bk) f(\br,\bk)\ ,
\end{equation}
where
\begin{equation}\label{W}
W^{ }_{\alpha_1\alpha_2}(\br,\bk) \!=\! \int d\br'
\phi^{ }_{\alpha_1}\left(\br \!+\! \frac{\br'}{2}\right)
e^{-\imath \bk \cdot \br'}
\phi^*_{\alpha_2}\left(\br \!-\! \frac{\br'}{2}\right)
\end{equation}
denotes the well-known Weyl-Wigner transform, and $\phi_\alpha(\br)$ the real-space wavefunction of the electronic state~$\alpha$.

Applying the inverse Weyl-Wigner transform (\ref{WWTinv}) to the original Wigner equation (\ref{WTE}), in the absence of external electromagnetic excitations the latter can be easily translated into the density-matrix equation
\begin{equation}\label{DME}
\frac{\partial \rho_{\alpha_1\alpha_2}}{\partial t}
=
\left.\frac{\partial \rho_{\alpha_1\alpha_2}}{\partial t}\right|_{\rm d}
+
\left.\frac{\partial \rho_{\alpha_1\alpha_2}}{\partial t}\right|_{\rm s}
\end{equation}
with
\begin{equation}\label{DMEd}
\left.\frac{\partial \rho_{\alpha_1\alpha_2}}{\partial t}\right|_{\rm d}
= 
\frac{\epsilon_{\alpha_1}-\epsilon_{\alpha_2}}{\imath\hbar}\,\rho_{\alpha_1\alpha_2}
\end{equation}
($\epsilon_{\alpha}$ denoting the energy of the single-particle state $\alpha$) and
\begin{equation}\label{DMEs}
\left.\frac{\partial \rho_{\alpha_1\alpha_2}}{\partial t}\right|_{\rm s}
=
S\left[\rho_{\alpha_1'\alpha_2'}\right]_{\alpha_1\alpha_2}\ ,
\end{equation}
where
\begin{equation}\label{Salpha}
S\left[\rho_{\alpha_1'\alpha_2'}\right]_{\alpha_1\alpha_2}
\equiv
\int \frac{d\br\,d\bk}{(2\pi)^3} W^{ }_{\alpha_1\alpha_2}(\br,\bk)
S\left[f(\br',\bk')\right](\br,\bk)
\end{equation}
and
\begin{equation}\label{WWT}
f(\br',\bk') =\! \sum_{\alpha_1'\alpha_2'} W_{\alpha_2'\alpha_1'}(\br',\bk') \rho_{\alpha_1'\alpha_2'}\, ={\rm tr}[\hat{W}(\br',\bk')\hat{\rho}]
\end{equation}
is the Weyl-Wigner transform previously recalled.
In the low-density limit considered above, the density-matrix version of the nonlocal scattering superoperator in (\ref{LDSM}) can be written as
\begin{equation}\label{DMEsLD}
\left.\frac{\partial \rho_{\alpha_1\alpha_2}}{\partial t}\right|_{\rm s}
=
\sum_{\alpha_1'\alpha_2'} A_{\alpha_1\alpha_2,\alpha_1'\alpha_2'} \rho_{\alpha_1'\alpha_2'}
+
B_{\alpha_1\alpha_2}\ ,
\end{equation}
where $A_{\alpha_1\alpha_2,\alpha_1'\alpha_2'}$ is given by
\begin{equation}\label{Aalpha}
\int \frac{d\br\,d\bk\;d\br'\,d\bk'}{(2\pi)^3}
W^{ }_{\alpha_1\alpha_2}(\br,\bk) 
A(\br,\bk;\br',\bk')
W^*_{\alpha'_1\alpha'_2}(\br',\bk')
\end{equation}
and
\begin{equation}\label{Balpha}
B_{\alpha_1\alpha_2}
=
\int \frac{d\br\,d\bk}{(2\pi)^3}
W^{ }_{\alpha_1\alpha_2}(\br,\bk) 
B(\br,\bk)\ .
\end{equation}
When the generic nonlocal scattering superoperator in (\ref{NLSS}) is approximated with the local form (\ref{LSS}), the related density-matrix version is still given by Eq.~(\ref{DMEs}), provided to replace Eq.~(\ref{Salpha}) with
\begin{equation}\label{Salphabis}
S\left[\rho_{\alpha_1'\alpha_2'}\right]_{\alpha_1\alpha_2}
\equiv
\int \frac{d\br\,d\bk}{(2\pi)^3} W^{ }_{\alpha_1\alpha_2}(\br,\bk)
\bar{S}\left[f(\br,\bk')\right](\br,\bk)\ .
\end{equation}
In particular, moving from the low-density nonlocal scattering superoperator (\ref{LDSM}) to its two local versions in (\ref{LDSMbis}) and (\ref{LDSMter}), their density-matrix counterparts are still given by Eq.~(\ref{DMEsLD}), replacing
the linear superoperator in (\ref{Aalpha}) with
\begin{equation}\label{Aalphabis}
\int \frac{d\br \; d\bk\,d\bk'}{(2\pi)^3}
W^{ }_{\alpha_1\alpha_2}(\br,\bk) 
\bar{A}(\br; \bk,\bk')
W^*_{\alpha'_1\alpha'_2}(\br,\bk')
\end{equation}
and
\begin{equation}\label{Aalphater}
\int \frac{d\br\,d\bk}{(2\pi)^3}
W^{ }_{\alpha_1\alpha_2}(\br,\bk) 
\bar{\bar{A}}(\br,\bk)
W^*_{\alpha'_1\alpha'_2}(\br,\bk)\quad,
\end{equation}
respectively.

While for the case of the generic nonlinear scattering superoperator in (\ref{DMEs}) it is extremely hard to draw conclusions about the fulfillment of the two basic requirements mentioned above (see Sec.~\ref{s-BTE} below), in the low-density limit one can relay on well-established criteria provided by the density-matrix theory applied to open quantum systems.\cite{b-Davies76,b-Breuer07}
More specifically, if (i) the linear superoperator $A_{\alpha_1\alpha_2,\alpha_1'\alpha_2'}$ in (\ref{DMEsLD}) is Lindblad-like,\cite{Lindblad76a} and (ii) the inhomogeneous term $B_{\alpha_1\alpha_2}$ is positive-definite, such scattering superoperator is known to preserve the positive-definite character of the density matrix $\rho_{\alpha_1\alpha_2}$.
It is however clear that the linear superoperator $A_{\alpha_1\alpha_2,\alpha_1'\alpha_2'}$ corresponding to a generic Wigner-function scattering kernel $A(\br,\bk; \br',\bk')$, and even more to the local versions in (\ref{Aalphabis}) and (\ref{Aalphater}) are not necessarily of Lindblad type, which implies that local Wigner-function scattering models may lead to positivity violations (see Secs.~\ref{s-LSM} and \ref{s-BTE} below).

From a physical point of view, the wide family of scattering superoperators (i.e., nonlinear versus linear ones as well as nonlocal versus local ones) examined so far can be divided into two main classes: carrier-nonconserving and carrier-conserving models.

Carrier-nonconserving models are often phenomenological in nature, and describe dissipation versus decoherence processes in terms of a few key macroscopic parameters; the prototypical example, examined in Sec.~\ref{s-RTA}, is the RTA model.

Conversely, carrier-conserving models are typically the result of a microscopic treatment of the interaction mechanism under investigation. In terms of the density-matrix formalism recalled so far, they are intrinsically trace-preserving, namely
\begin{equation}\label{TP1}
\sum_\alpha S\left[\rho_{\alpha_1'\alpha_2'}\right]_{\alpha\alpha} = 0\ .
\end{equation}
In the low-density limit such carrier-conserving scattering approaches reduce to a trace-preserving linear superoperator (the inhomogeneous term $B_{\alpha_1\alpha_2}$ in (\ref{DMEsLD}) being absent), namely
\begin{equation}\label{DMEsLDbis}
\left.\frac{\partial \rho_{\alpha_1\alpha_2}}{\partial t}\right|_{\rm s}
=
\sum_{\alpha_1'\alpha_2'} A_{\alpha_1\alpha_2,\alpha_1'\alpha_2'} \rho_{\alpha_1'\alpha_2'}
\end{equation}
with
\begin{equation}\label{TP2}
\sum_\alpha A_{\alpha\alpha,\alpha_1'\alpha_2'} = 0\ .
\end{equation}
The most popular example of carrier-conserving scattering model, examined in Sec.~\ref{s-BTE}, is the well-known Boltzmann collision term of the semiclassical transport theory.

\section{Local Scattering Models}\label{s-LSM}
In this section we point out that, under some circumstances, the assumption (\ref{LSS}) of space locality for the scattering superoperator may give rise to pathological behaviors.

\subsection{The RTA model}\label{s-RTA}
As pointed out in Sec.~\ref{s-WFDM}, the RTA model, described by Eq.~(\ref{RTA}), is a particular case of the fully local scattering superoperator in (\ref{LDSMter}).
Thanks to its intuitive simplicity and easiness of implementation, the RTA model has been widely applied to a large variety of problems involving electronic dissipation versus decoherence in nanodevices, such as the analysis of current-voltage characteristics and the carrier density profile in resonant tunneling semiconductor heterostructures and superlattices.\cite{b-Jacoboni89,b-Rossi11,b-Jacoboni10}

Notably, RTA schemes have been applied both in semiclassical-transport simulations\cite{b-Jacoboni89} and in the quantum regime through, e.g., Wigner-function treatments.\cite{Frensley90a} 
However, while for the semiclassical transport theory the range of validity of such approximation is well established, much less is known about its soundness for quantum-transport simulations based on the Wigner-function formalism.\\
It is easy to show that in this case the corresponding density-matrix version in (\ref{DMEsLD}) reduces to
\begin{equation}\label{RTAbis}
\left.\frac{\partial \rho_{\alpha_1\alpha_2}}{\partial t}\right|_{\rm s}
=
-\sum_{\alpha'_1\alpha'_2}
\Gamma_{\alpha_1\alpha_2,\alpha'_1\alpha'_2}
\left(\rho^{ }_{\alpha'_1\alpha'_2} - \rho^\circ_{\alpha'_1\alpha'_2}\right)
\end{equation}
with
\begin{equation}\label{Gamma}
\Gamma_{\alpha_1\alpha_2,\alpha'_1\alpha'_2}
=
\int \frac{d\br\,d\bk}{(2\pi)^3}
W^{ }_{\alpha_1\alpha_2}(\br,\bk) 
\Gamma(\br,\bk)
W^*_{\alpha'_1\alpha'_2}(\br,\bk)\ ,
\end{equation}
where $\rho^\circ_{\alpha_1\alpha_2} = f^\circ_{\alpha_1} \delta_{\alpha_1\alpha_2}$ denotes the (diagonal) equilibrium density matrix. 

It is now worth comparing Eq.~(\ref{RTAbis}) (obtained by applying a semiclassical RTA model to the Wigner function) to the equation obtained applying the RTA scheme straightforwardly to the density-matrix evolution, i.e.,
\begin{equation}\label{RTADME}
\left.\frac{\partial \rho_{\alpha_1\alpha_2}}{\partial t}\right|_{\rm s}
=
-\,\frac{\Gamma_{\alpha_1} + \Gamma_{\alpha_2}}{2}\,\left(\rho^{ }_{\alpha_1\alpha_2} - \rho^\circ_{\alpha_1\alpha_2}\right)\ ,
\end{equation}
where $\Gamma_\alpha$ can be regarded as a sort of state-dependent effective inverse life-time, involving all relevant interaction mechanisms acting on the carrier in state $\alpha$; indeed, in the semiclassical limit ($\rho_{\alpha_1\alpha_2} = f_{\alpha_1} \delta_{\alpha_1\alpha_2}$) the density-matrix RTA model (\ref{RTADME}) reduces to the well-known injection-loss scheme\cite{Iotti05b} of conventional device modelling: \footnote{
In addition to the relaxation dynamics (\ref{RTADMEf}) of the level population $f_\alpha$, the density-matrix RTA model in (\ref{RTADME}) describes the scattering-induced decay of the interlevel polarization known as decoherence process.\cite{Rossi02b}}
\begin{equation}\label{RTADMEf}
\left.\frac{\partial f_\alpha}{\partial t}\right|_{\rm s}
=
- \Gamma_\alpha \left(f^{ }_\alpha - f^\circ_\alpha\right)
= S_\alpha - \Gamma_\alpha f_\alpha\ .
\end{equation}
As one can see, while the scattering superoperator in (\ref{RTADME}) is diagonal in the single-particle
basis $\alpha$, Eq.~(\ref{RTAbis}) is non-diagonal in such basis. As a consequence, it does not necessarily preserve the positive-definite nature of the single-particle density matrix (see Figs.~\ref{Fig2} and \ref{Fig3} below), a basic physical prerequisite of any quantum-transport simulation scheme.
The only exception is the case of a space- and energy-independent relaxation rate, $\Gamma(\br,\bk) = \Gamma_0$, as well as a state-independent rate, $\Gamma_\alpha = \Gamma_0$, for which the two RTA models in (\ref{RTAbis}) and (\ref{RTADME}) coincide.

\begin{figure}
\centering
\includegraphics[width=\columnwidth]{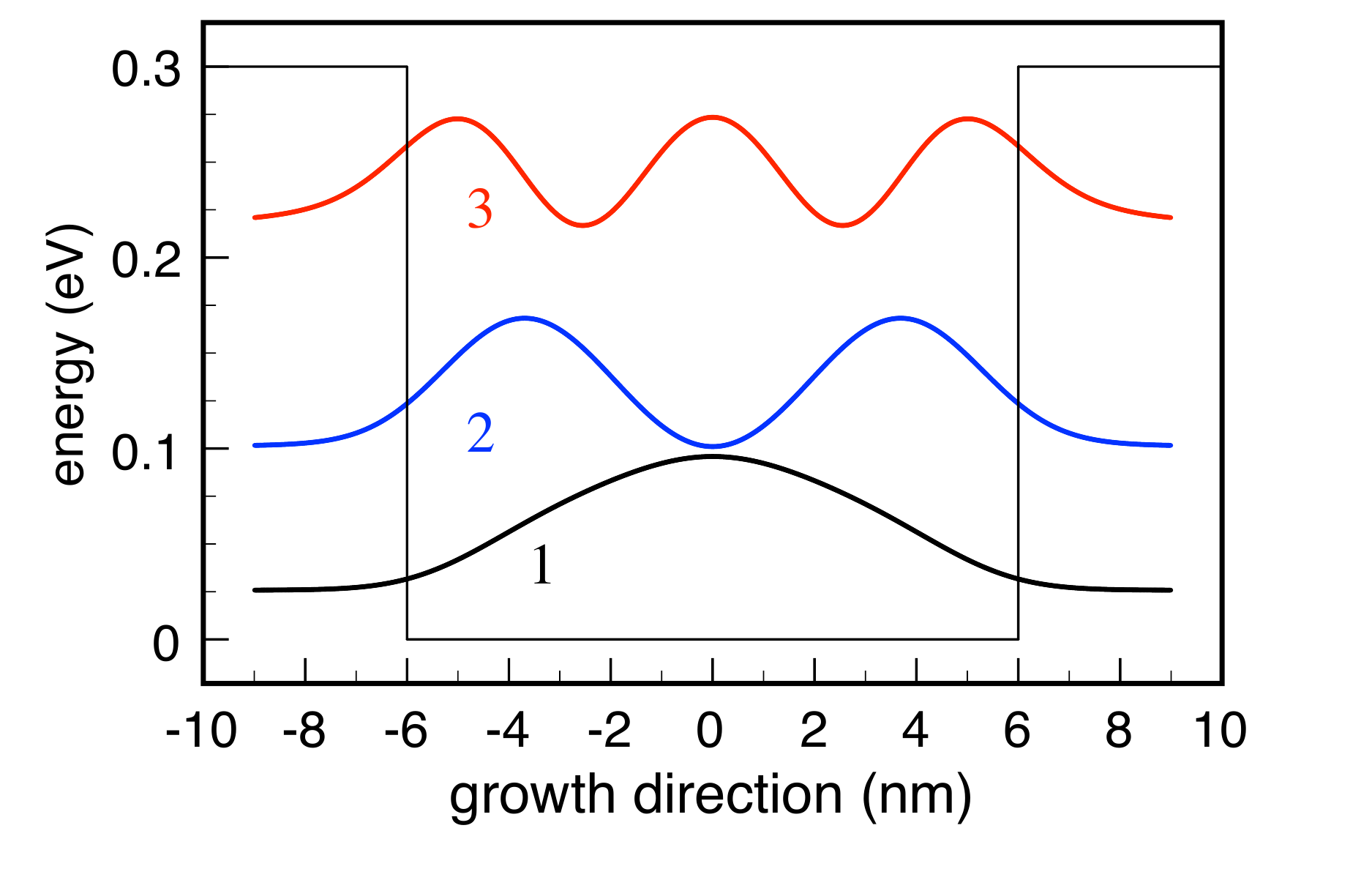}
\caption{ 
Conduction band profile along the growth direction for the prototypical GaAs/(Al,Ga)As 
QW nanostructure considered in our simulated experiments. Energy levels of the three 1D bound states are shown, together with the corresponding squared wavefunctions.}
\label{Fig1}
\end{figure}

In order to point out intrinsic limitations of the conventional RTA model (\ref{RTA}) within the Wigner-function picture (or its equivalent density-matrix formulation in (\ref{RTAbis})), let us consider the basic nanosystem depicted in Fig.~\ref{Fig1}. It consists of a $l = 12$\,nm thick GaAs quantum well (QW) surrounded by (Al,Ga)As barriers with band offset $V_\circ = 0.3$\,eV; its three-dimensional (3D) electronic states exhibit the usual subband structure due to confinement along the growth direction, $z$.
To simplify our analysis, in the remainder of this section we shall neglect in-plane phase-space coordinates and adopt an effective one-dimensional (1D) description of the QW nanosystem, i.e., $(\br,\bk) \equiv (z,k)$. This implies that, within such simplified treatment, the set of single-particle quantum numbers of our nanostructure coincides with the partially discrete index of our 1D states only: $\alpha \equiv {\rm n}$.
Moreover, for all the low-temperature simulated experiments discussed below we shall consider as initial condition the first excited state (${\rm n} = 2$) of the QW, while the ground state (${\rm n} = 1$) corresponds to the thermal-equilibrium condition.

More specifically, we shall present and compare simulated experiments for three different RTA models: the model referred to as M1 assumes a space- and energy-independent rate $\Gamma_0$ and is often called phenomenological or macroscopic RTA approach.
The model referred to as M2 employs a space-dependent but energy-independent $\Gamma(z)$, which is taken vanishing within the interval $-\frac{a}{2} < z < +\frac{a}{2}$, and equal to $\Gamma_0$ outside.
Finally, the model referred to as M3 considers a space-independent but momentum/energy-  dependent rate $\Gamma(k)$, which is vanishing for $|k| < k_{\rm min} = \sqrt{2 m^* \epsilon_{\rm min}}/\hbar$ and equal to~$\Gamma_0$ otherwise.
In particular, model M2 may partially mimic what happens in a complex multi-layer quantum structure, whose semiclassical scattering rates are typically material-dependent; in a similar way, model M3 describes the conventional energy-threshold scenario typical of electron-optical phonon scattering in a variety of nanomaterials and related nanodevices, including, e.g., new-generation quantum-cascade lasers.\cite{Iotti05b}

\begin{figure}
\centering
\includegraphics[width=\columnwidth]{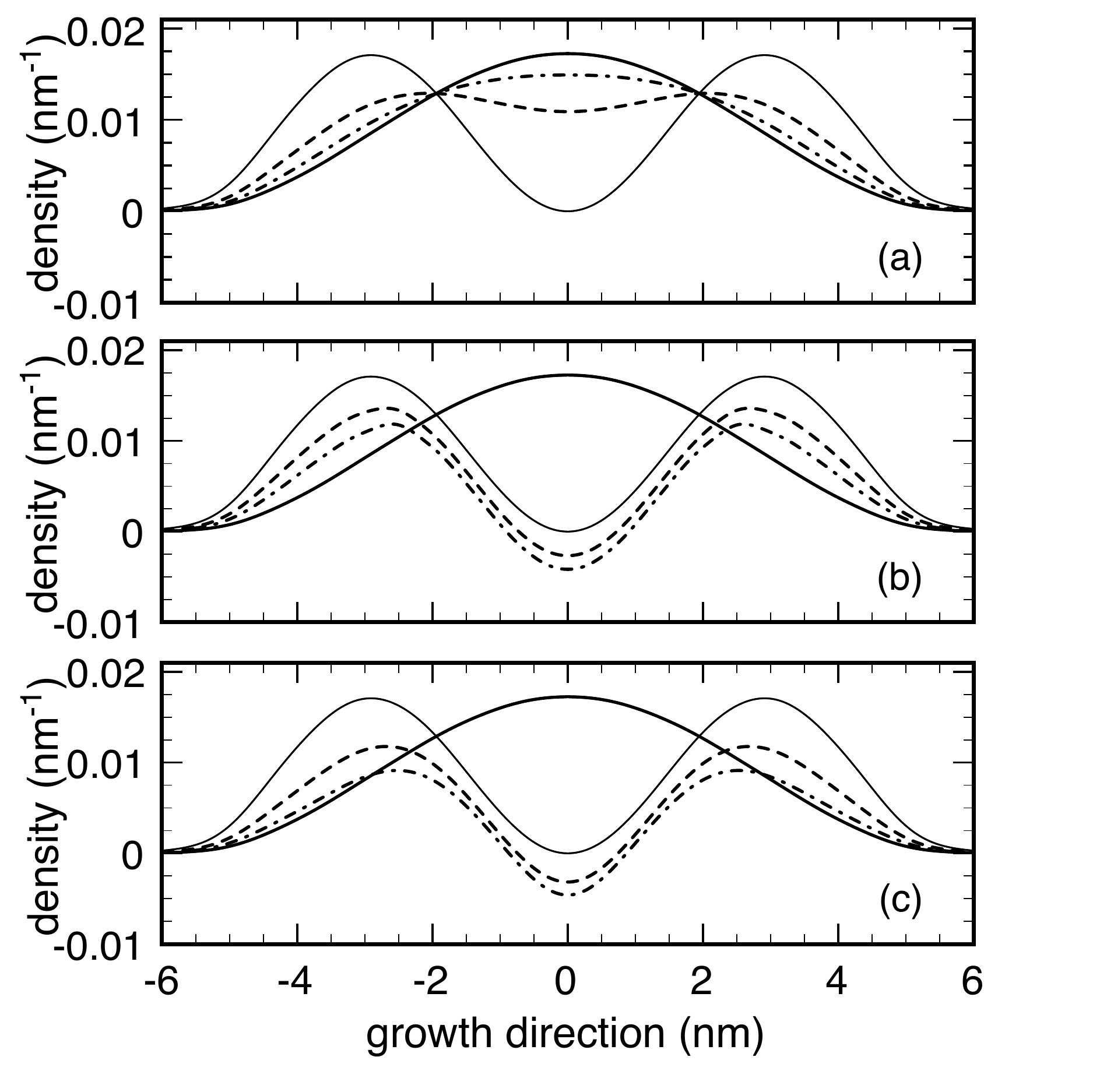}
\caption{ 
Low-temperature dissipation dynamics in the $l = 12$\,nm thick GaAs/(Al,Ga)As QW of Fig.~\ref{Fig1}. The carrier density $n(z)$ in (\ref{nz}) is plotted as a function of position at different times ($0$\,ps: thin solid curve, $0.8$\,ps: dashed curve, and $1.6$\,ps: dash-dotted curve) for models M1 (a), M2 (b), and M3 (c). The following parameters are employed: $1/\Gamma_0 = 0.8$\,ps for M1,
$1/\Gamma_0 = 0.4$\,ps
and $a = 5$\,nm for M2, and $1/\Gamma_0 = 0.5$\,ps
and $\epsilon_{\rm min} = 40$\,meV for M3.
The equilibrium carrier density is also shown for comparison (solid curve).
}
\label{Fig2}
\end{figure}

Figure \ref{Fig2} shows a comparison between the different dissipation dynamics induced on the prototypical QW system of Fig.~\ref{Fig1} by the three RTA models just described. More specifically, here we show the spatial carrier-density profile along the growth direction ($z$) corresponding to the 1D Wigner function $f(z,k)$, namely
\begin{equation}\label{nz}
n(z) = \frac{1}{2 \pi}\,\int dk\,f(z,k)\ ,
\end{equation} 
at different times, obtained solving the 1D version ($\br,\bk \equiv z,k$) of the Wigner transport equation (\ref{WTE}) equipped with the quantum-mechanical deterministic term in (\ref{WTEd}) as well as with the conventional RTA scheme in (\ref{RTA}).

The relaxation dynamics resulting from M1 [panel (a)] exhibits a well-established and physically sound scenario: the initial charge distribution corresponding to the QW first excited state ${\rm n} = 2$ (thin solid curve) decays exponentially, and at the same time one observes the progressive population of the QW ground state ${\rm n} = 1$; as a result, after $1.6$\,ps (dash-dotted curve) the state of the electronic system is not too far from its equilibrium carrier distribution (solid curve).

The scenario is substantially different in the case of both a space-dependent rate [M2, panel (b)] and an energy-dependent one [M3, panel (c)]. 
One observes, in particular, a significant slowdown of the excited-level decay process, and after $1.6$\,ps (dash-dotted curve) the charge distribution is still far from its equilibrium counterpart (solid curve); in addition to such unexpected behaviour, negative carrier distributions arise in both models. This is an unambiguous fingerprint of unphysical electronic states, which emphasizes the difference with the results displayed in panel (a).

\begin{figure}
\centering
\includegraphics[width=\columnwidth]{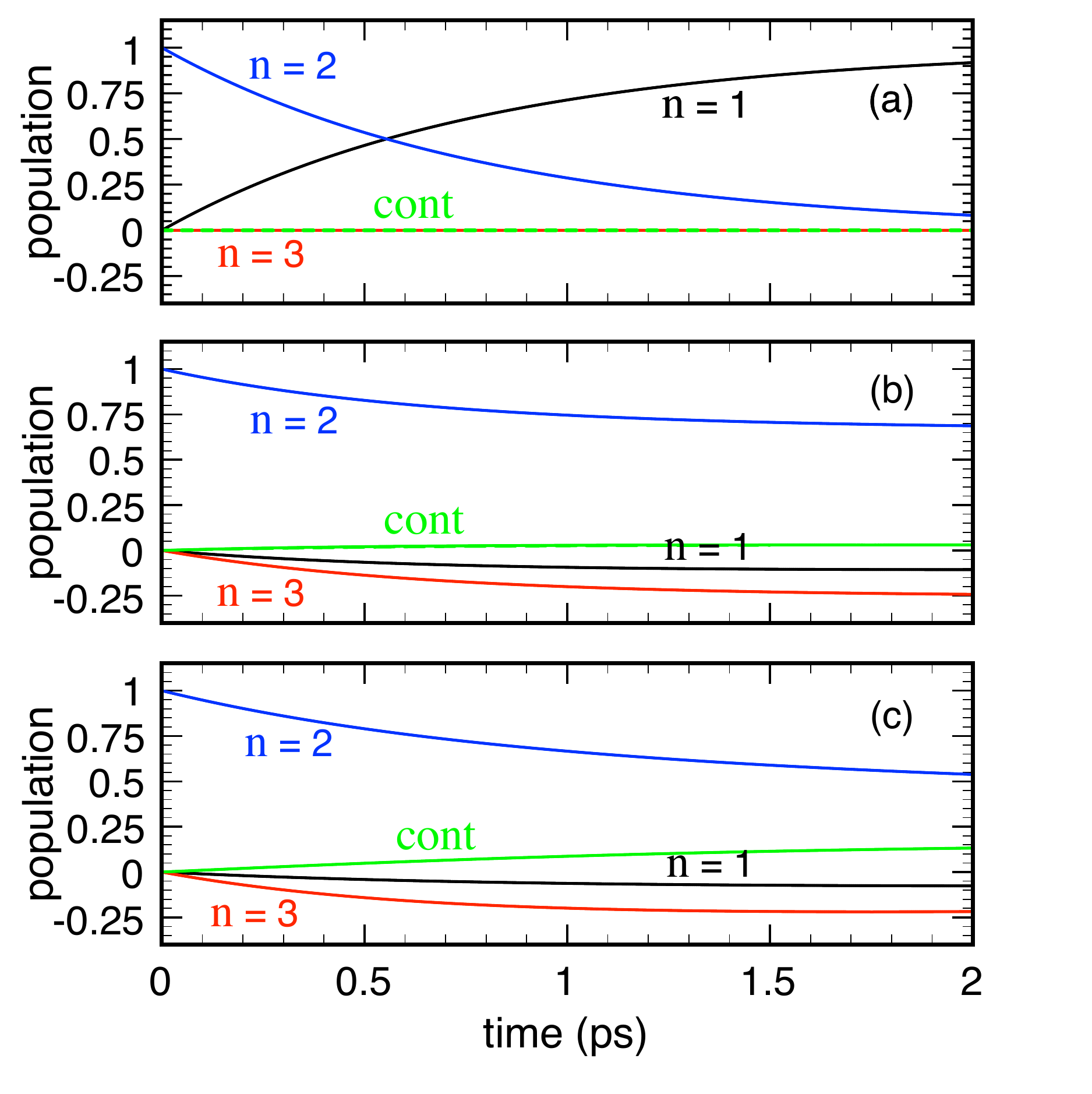}
\caption{(Color online) 
Bound-state populations [see Eq.~(\ref{frmn})] as well as the continuum (cont) contribution as a function of time for models M1 (a), M2 (b), and M3 (c).
}
\label{Fig3}
\end{figure}

To better identify the origin of these anomalous relaxation profiles in real space, we have evaluated the carrier populations $f_\alpha \equiv \rho_{\alpha\alpha}$ by means of the inverse Weyl-Wigner transform (\ref{WWTinv}) ; more specifically, for our 1D model ($\alpha \equiv {\rm n}$) we have:
\begin{equation}\label{frmn}
f_{\rm n} = \int \frac{dz\,dk}{2\pi} W^{ }_{{\rm n}{\rm n}}(z,k) f(z,k)\ .
\end{equation}
Figure \ref{Fig3} shows the time evolution of the populations of the three QW bound states ($f_1$, $f_2$, and $f_3$) as well as of the continuum one, defined as
$N_{\rm cont} = N_{\rm tot} - \sum_{{\rm n} = 1}^{3} f_{\rm n}$ 
with $N_{\rm tot}$ denoting the total number of carriers.

As expected, for model M1 [see panel (a)] we observe a simple exponential depopulation of the initial level ${\rm n} =2$, accompanied by a corresponding population of the ground state ${\rm n} = 1$, while the populations of level ${\rm n} = 3$ and of the continuum are not affected by the relaxation dynamics.
Once again, moving to models M2 [panel (b)] and M3 [panel (c)], the physically sound scenario of model M1 is lost: for both M2 and M3 we do observe (i) a significant slowdown in the depopulation of the initial state ${\rm n} = 2$, (ii) negative population values for both state ${\rm n} = 1$ and ${\rm n} = 3$, and (iii) a notable population of the continuum, particularly for model M3, in spite of our zero-temperature analysis.

The level-population analysis in Fig.~\ref{Fig3}
clearly shows that the conventional Wigner-function RTA term (\ref{RTA}) (or its equivalent density-matrix formulation in (\ref{RTAbis})) does not preserve the positivity of the density matrix $\rho_{\alpha_1\alpha_2}$, and induces a fictitious interlevel coupling; the latter, in turn, may also lead to an artificial generation of interlevel phase coherence, additional fingerprint of an unphysical dissipation dynamics.
Moreover, in spite of the fact that both model (\ref{RTAbis}) and (\ref{RTADME}) share the correct steady-state solution $\rho^\circ_{\alpha_1\alpha_2}$, the spectrum of the (non-diagonal) superoperator (\ref{Gamma}) may involve eigenvalues with negative real parts, leading to pathological divergences in the system dynamics; this is similar to the case of Lindblad versus non-Lindblad Markov models.\cite{Taj09b,Rosati14e}\\

It is finally worth noticing that the RTA scheme in~(\ref{RTAbis}) is not trace-preserving, which implies that the total amount of charge in the device {\it is not conserved}. Indeed the sum of the various level populations (three bound states plus continuum) shown in the middle and lower panels of Fig.~\ref{Fig3} slightly changes with time.
This is an intrinsic feature of any type of RTA approach, which is well known to arise in the semiclassical transport theory as well.

\subsection{Boltzmann-like scattering model}
\label{s-BTE}
One may at first think that the unphysical behaviours pointed out in Sec.~\ref{s-RTA} originate from the lack of trace conservation characterizing the RTA scheme, and/or from  the effective 1D modelling of the QW  in Fig.~\ref{Fig1}. This is, however, not the case. 
Here below we shall argue that similar problems arise when:  
(i) the RTA scheme in Eq.~(\ref{RTA}) is replaced by the (trace-preserving) Boltzmann collision term in Eq.~(\ref{BTE1}); (ii) a fully 3D treatment of the prototypical QW nanostructure is adopted.

The Boltzmann collision term (\ref{BTE1}) is characterized by the well-established in- minus out-scattering structure; indeed, the latter may also be written as
\begin{eqnarray}\label{BTE3}
\left.\frac{\partial f(\br,\bk)}{\partial t}\right|_{\rm s}
&=&
\int d\br'\,d\bk'
P^{\rm in}(\br,\bk;\br',\bk') f(\br',\bk')
\nonumber \\
&-&
\int d\br'\,d\bk'
P^{\rm out}(\br,\bk;\br',\bk') f(\br',\bk')
\end{eqnarray}
with
\begin{equation}\label{Pin}
P^{\rm in}(\br,\bk;\br',\bk') 
=
\delta(\br-\br') P(\br;\bk,\bk')
\end{equation}
and
\begin{equation}\label{Pout}
P^{\rm out}(\br,\bk;\br',\bk') 
=
\delta(\br-\br') \delta(\bk-\bk')
\int d\bk'' P(\br;\bk'',\bk)\ ,
\end{equation}
which confirms that both superoperators are local in $\br$, and that the out-scattering one is local in $\bk$ as well.

In order to provide a microscopic (i.e., parameter-free) description of energy dissipation, we have replaced the partially phenomenological 1D modelling of Sec.~\ref{s-RTA} with a fully microscopic 3D treatment; 
more specifically, (i) the partially discrete 1D energy spectrum ($\alpha \equiv {\rm n}$) of the QW nanosystem has been replaced by its fully 3D subband structure ($\alpha \equiv \bk_\parallel {\rm n}$), and 
(ii) the 3D scattering rates in (\ref{BTE2}) have been derived via the conventional Fermi's golden rule assuming as main dissipation source carrier-LO phonon interaction within a GaAs bulk crystal.\cite{b-Jacoboni89} As a result, the latter are space-independent.
Within such 3D description the Weyl-Wigner phase-space of the QW nanosystem is given by $\br \equiv \br_\parallel,z$ and $\bk \equiv \bk_\parallel,k_z$.
This implies that the effective 1D carrier-density profile along the growth direction ($z$) in (\ref{nz}) is now replaced by
\begin{equation}\label{nzbis}
n(z) = \frac{1}{(2 \pi)^3}\,\int d\br_\parallel\,d\bk\,f(\br,\bk)\ ,
\end{equation} 
and the effective 1D level population in (\ref{frmn}) is replaced by the following subband population:
\begin{equation}\label{frmnbis}
f_{\rm n} = 
\sum_{{\rm k}_\parallel}
\int \frac{d\br\,d\bk}{(2\pi)^3}\, 
W^{ }_{{\rm k}_\parallel {\rm n},{\rm k}_\parallel {\rm n}}(\br,\bk) f(\br,\bk)\ .
\end{equation}

\begin{figure}
\centering
\includegraphics[width=\columnwidth]{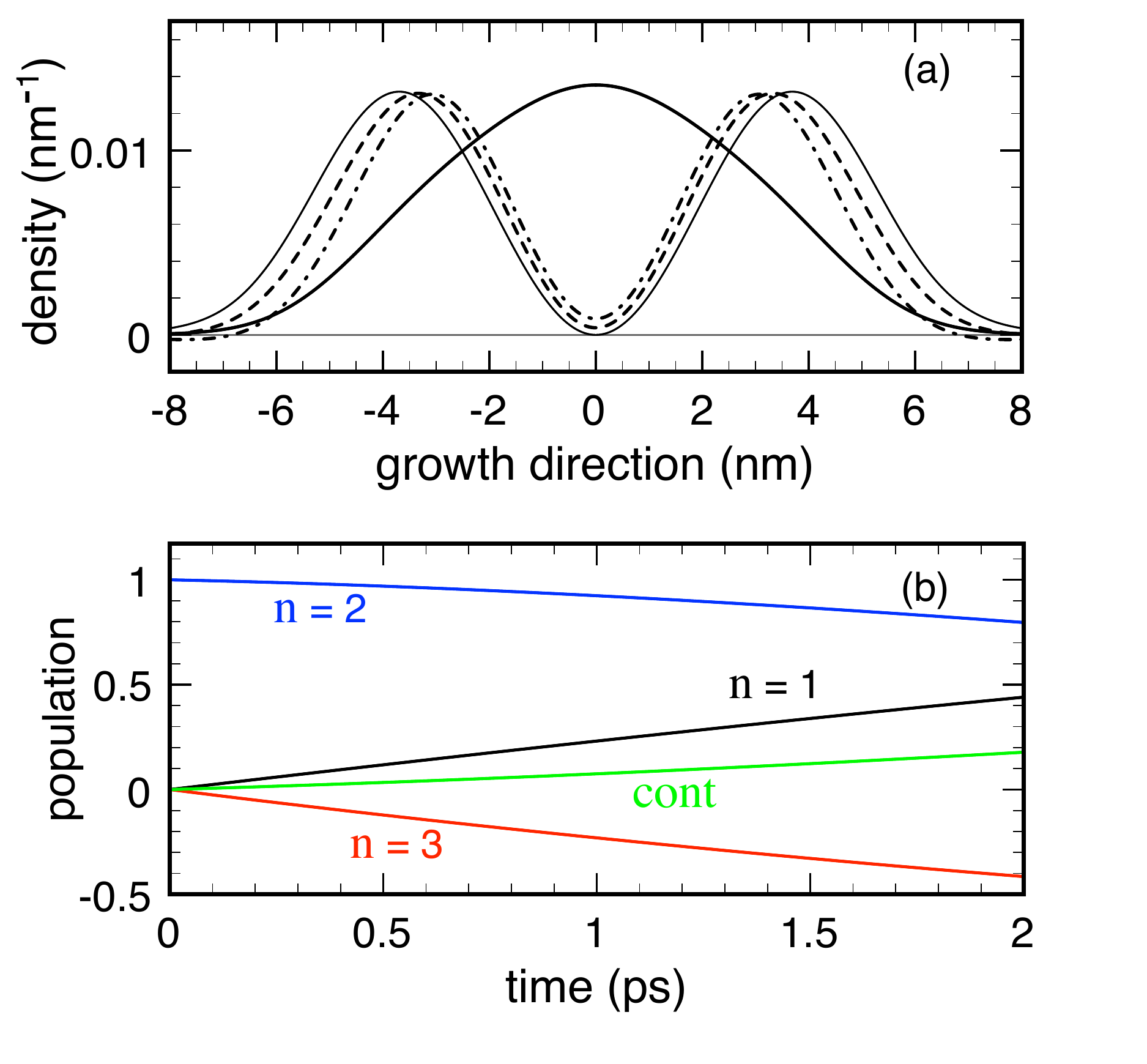}
\caption{(Color online)
Low-temperature dissipation dynamics in the $l = 12$\,nm thick GaAs/(Al,Ga)As QW of Fig.~\ref{Fig1} resulting from the fully 3D local Boltzmann model in (\ref{BTE1}) in the low-density limit. 
(a) Carrier density $n(z)$ in (\ref{nzbis}) as a function of position at different times ($0$\,ps -- thin solid curve, $0.8$\,ps -- dashed curve, and $1.6$\,ps -- dash-dotted curve); here, the equilibrium carrier density is also shown for comparison (solid curve). 
(b) Bound-state subband populations [see Eq.~(\ref{frmnbis})] as well as continuum (cont) contribution as a function of time.
}
\label{Fig4}
\end{figure}

Figure \ref{Fig4} shows the
low-temperature and low-density energy-dissipation dynamics in the GaAs-based QW nanostructure (see Fig.~\ref{Fig1}) resulting from the fully 3D Boltzmann bulk model just described.
Compared to the unphysical results obtained via the RTA model [see panels (b) and (c) in Figs.~\ref{Fig2} and \ref{Fig3}], the spatial carrier density (panel a) obtained via the Boltzmann scattering model is less affected by negative-value regions, but, exactly as for the RTA case, one observes again a significant slowdown of the excited-level decay process, and after $1.6$\,ps (dash-dotted curve) the charge distribution is still far from its equilibrium counterpart (solid curve).
Such anomalous scenario is fully confirmed by the subband-population analysis (panel b) which shows (i) a significant depopulation slowdown of the initial subband ${\rm n} = 2$, (ii) negative population values for subband ${\rm n} = 3$, and (iii) in spite of our zero-temperature analysis, a relevant (negative) population of the continuum.

\begin{figure}
\centering
\includegraphics[width=\columnwidth]{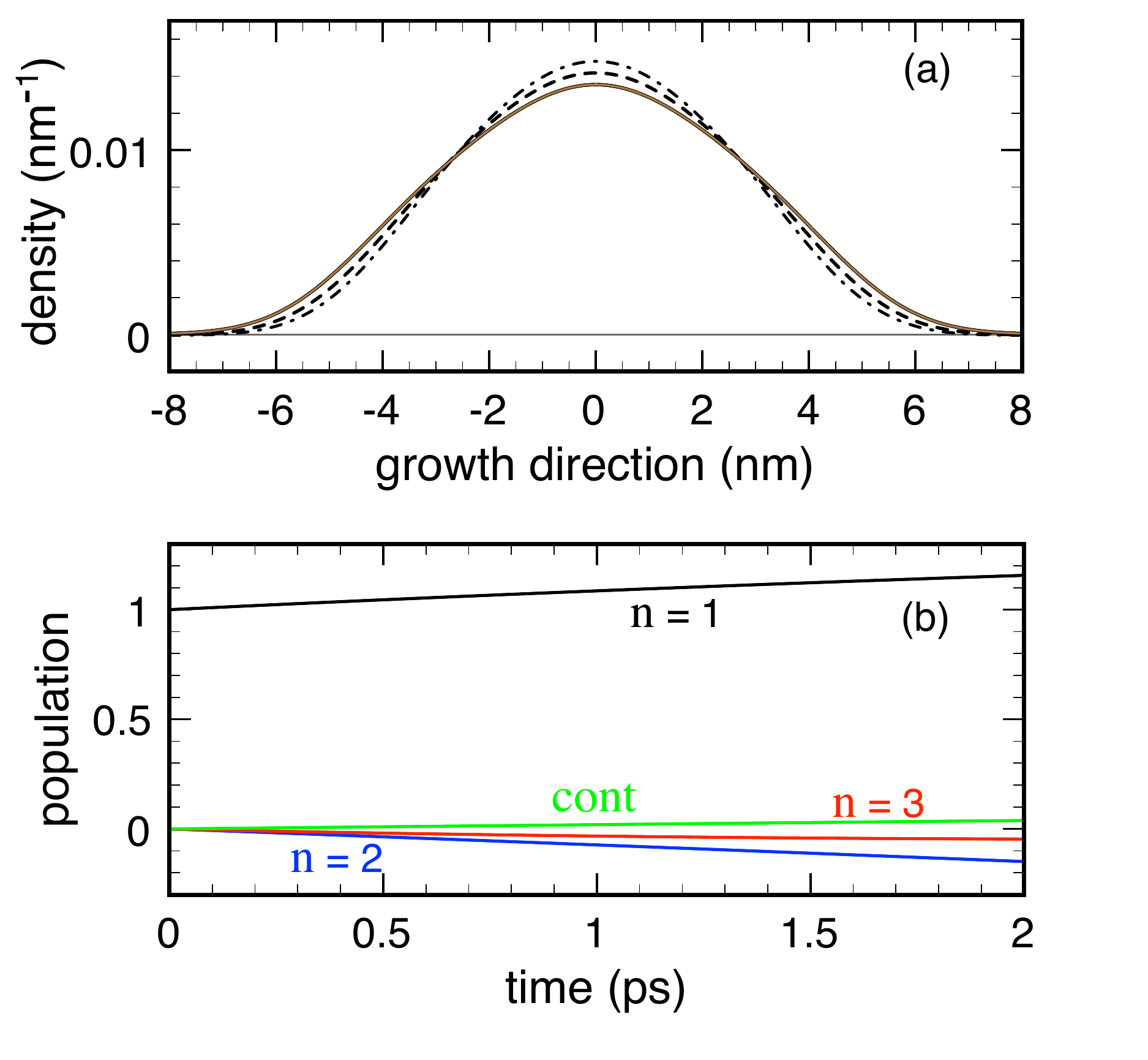}
\caption{(Color online)
Same as in Fig.~\ref{Fig4} but considering as initial state the thermal-equilibrium state ${\rm n} = 1$ (see text).
}
\label{Fig5}
\end{figure}

To deepen our analysis, we have repeated the simulated experiment on the GaAs/(Al,Ga)As QW nanosystem replacing the initial condition employed so far (i.e., the excited state ${\rm n} = 2$) with the QW thermal-equilibrium state ${\rm n} = 1$.
The new results, reported in Fig.~\ref{Fig5}, are rather counterintuitive and confirm again intrinsic limitations of such local treatments.
Indeed, in this case a correct scattering superoperator is expected to leave the electronic system in its equilibrium state. However, Fig.~\ref{Fig5} shows that the Boltzmann collision term in (\ref{BTE1}) drives the electronic system out of equilibrium, giving rise to a steady-state solution characterized again by negative populations and by a significant population of the continuum, in striking contrast with the low-temperature regime.

The results reported in Figs.~\ref{Fig4} and \ref{Fig5} thus clearly show that the pathological behaviors obtained via the RTA model (see Sec.~\ref{s-RTA}) may also affect Boltzmann-like scattering superoperators. In particular, the key feature shared by both models is their local character, which is known to be intrinsically incompatible with any quantum-mechanical treatment.

More specifically, the physical origin of the two main anomalous behaviors pointed out so far, namely the dissipation slowdown and the wrong thermalization dynamics, can be explained as follows.
Both the RTA simulated experiments in Figs.~\ref{Fig2}-\ref{Fig3} and the Boltzmann-like ones in Figs.~\ref{Fig4}-\ref{Fig5} are based on bulk-like scattering models; indeed, both the relaxation-time $\Gamma(\br,\bk)$ in (\ref{RTA}) and the semiclassical scattering rate $P(\br;\bk,\bk')$ in (\ref{BTE1}) refer to a semiconductor bulk crystal, i.e., they do not account for the electronic subband structure of the nanosystem. In particular, the scattering rates entering the Boltzmann collision term in (\ref{BTE1}) are evaluated via the conventional Fermi's golden rule using as noninteracting states standard 3D plane waves, instead of the nanostructure single-particle wavefunctions $\phi_\alpha(\br)$.
It follows that the two terms entering the Wigner transport equation (\ref{WTE}) are intrinsically incompatible: while the deterministic one in (\ref{WTEd}) accounts for the QW subband structure via the nanomaterial confinement potential $V(\br)$, the same does not apply to the scattering term.
Such a significant description mismatch, recently pointed out also within the density-matrix formalism,\cite{Zhan16a} is responsible for the dissipation slowdown previously mentioned as well as for the wrong thermalization dynamics reported in Fig.~\ref{Fig5}. Indeed, the steady-state solution of the bulk-like Boltzmann collision term in (\ref{BTE1}) is simply given by the (space-independent) Fermi-Dirac distribution. The latter has nothing to do with the thermal-equilibrium Wigner function $f^\circ(\br,\bk)$ of the QW nanosystem, which is always space-dependent and significantly different from zero within the QW region only.

\section{Proposed nonlocal scattering models}
\label{s-NLA}
\subsection{Nonlocal RTA model}
In order to overcome the serious limitations of the local models pointed out in Sec.~\ref{s-LSM}, we first propose an alternative RTA scheme for the Wigner function. Our strategy is to start from the density-matrix RTA model in (\ref{RTADME}) and apply to it the Weyl-Wigner transform  (\ref{WWT}) as well as its inverse in (\ref{WWTinv}), therefore obtaining 
\begin{equation}\label{RTAnew}
\left.\frac{\partial f(\br,\bk)}{\partial t}\right|_{\rm s}
\!=\!
- \int d\br'\,d\bk'\, \Gamma(\br\!,\!\bk;\br'\!,\!\bk') \left(\!f(\br',\bk') \!-\! f^\circ(\br',\bk')\right)
\end{equation}
where
\begin{equation}\label{GammaNL}
\Gamma(\br\!,\!\bk;\br'\!,\!\bk')�
=
\sum_{\alpha_1\alpha_2}
W^{ }_{\alpha_1\alpha_2}(\br,\bk)
\,\frac{\Gamma_{\alpha_1} + \Gamma_{\alpha_2}}{16\pi^3}\,
W^*_{\alpha_1\alpha_2}(\br',\bk')
\end{equation}
is a fully nonlocal RTA superoperator expressed in terms of the relaxation rates $\Gamma_\alpha$ of the density-matrix theory.
In striking contrast to the standard RTA model in (\ref{RTA}), the generalized version in (\ref{RTAnew})   intrinsically ensures the positivity of the spatial charge density $n(z)$ as well as of the level populations $f_{\rm n}$. 
Indeed, (i) as shown in Ref.~[\onlinecite{Dolcini13a}], the RTA model in (\ref{RTADME}) is known to preserve the positive-definite character of the density matrix $\rho_{\alpha_1\alpha_2}$, and (ii) the action of the Weyl-Wigner transform in (\ref{WWT}) does not alter such property.

\begin{figure}
\centering
\includegraphics[width=\columnwidth]{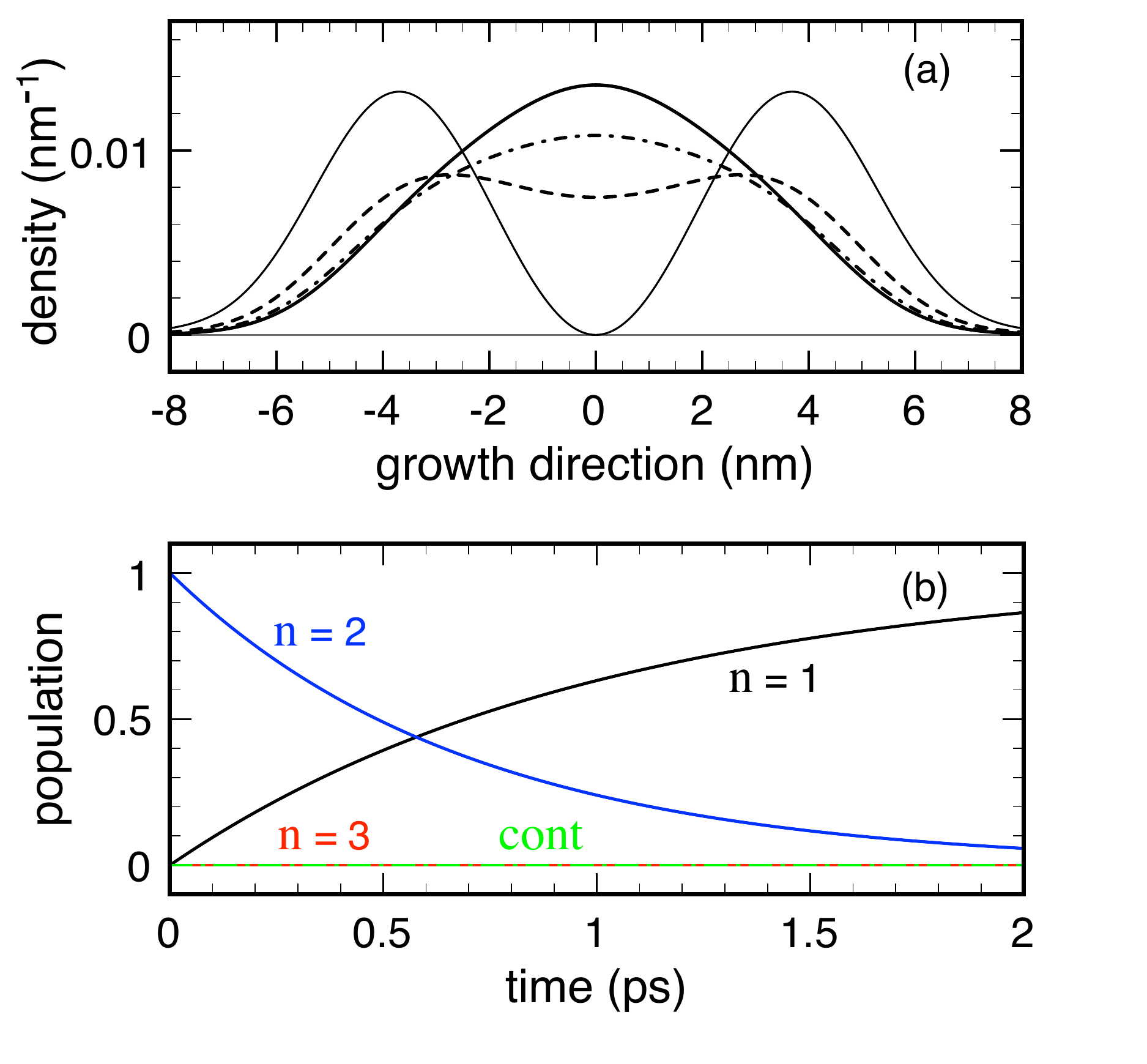}
\caption{(Color online) 
Same as in Fig.~\ref{Fig2} (panel a) and in Fig.~\ref{Fig3} (panel b) but for the proposed nonlocal RTA model in (\ref{RTAnew}) with $1/\Gamma_1 = 1$\,ps and $1/\Gamma_2 = 0.7$\,ps.
}
\label{Fig6}
\end{figure}

To illustrate the effects of our result, we apply the proposed nonlocal RTA model (\ref{RTAnew}) to the prototypical QW nanosystem in Fig.~\ref{Fig1} adopting again the very same effective 1D model described above.  
Figure \ref{Fig6} shows the results obtained by setting $1/\Gamma_1 = 1$\,ps and $1/\Gamma_2 = 0.7$\,ps.
As one can see, opposite to the pathological behaviors obtained via the RTA models M2 and M3 (see panels b and c in Figs.~\ref{Fig2} and \ref{Fig3}), here both the spatial carrier distributions (panel a) and the corresponding level populations (panel b) are always positive-definite. Moreover, as for the case of the physically sound results of model M1, we deal again with a correct relaxation dynamics toward the equilibrium state (solid curve in panel a) without unphysical interlevel relaxation couplings and continuum population.\\

We conclude this subsection by commenting about the range of validity of the proposed nonlocal RTA model. At a semiclassical level, the RTA model is known to properly describe a few scattering mechanisms only, namely elastic or quasi-elastic, as well as inelastic isotropic processes,\cite{b-Jacoboni89} in quasi-equilibrium and low-density conditions. These limitations apply to the conventional Wigner-function RTA modeling discussed in Sec.~\ref{s-RTA} as well as to its nonlocal generalization in (\ref{RTAnew}). To overcome such limitations, the key step is to replace the RTA modeling with Boltzmann-like treatments (see below).\\
Furthermore, the RTA scheme in Eq.(\ref{RTAnew}) is not trace-preserving, as can be straightforwardly seen from the density-matrix RTA model, Eq.(\ref{RTADME}), it originates from. This implies that the total amount of charge in the device is not conserved. Indeed the sum of the various level populations (three bound states plus continuum) shown in  Fig.~\ref{Fig6} slightly changes with time.
The nonlocal RTA scheme can thus be applied as long as the total population variation is small compared to the initial condition. We stress once again that this is an intrinsic feature of any type of RTA approach (see also Fig.~\ref{Fig3}), which also arises in the semiclassical transport theory and is by no means specific of the nonlocal RTA scheme proposed here.
In order to remove such constraint, it is imperative to replace the RTA models considered so far with genuine trace-preserving scattering superoperators.

\subsection{Nonlocal Boltzmann-like scattering model}
Importantly, the generalization scheme employed to derive the nonlocal RTA model in (\ref{RTAnew}) can also be adopted to overcome the  limitations of the Boltzmann-like   treatment pointed out in Sec.~\ref{s-BTE}. Instead of defining the scattering superoperator directly within the Weyl-Wigner phase-space ($\br,\bk$) one can start from a reliable dissipation model for the density matrix, and translate it into the Wigner-function picture.
To this aim, our starting point is the nonlinear density-matrix treatment recently proposed in Ref.~[\onlinecite{Rosati14e}]; indeed, the latter
(i) applies to any generic nanostructure, (ii) accounts for high-density effects, (iii) provides the correct thermal-equilibrium state, and (iv) preserves the positive-definite character of the single-particle density matrix.
More specifically, for both carrier-phonon and carrier-carrier interaction mechanisms, energy dissipation and decoherence is described by the nonlinear scattering superoperator
\begin{widetext}
\begin{equation}\label{DMEsbis}
\left.\frac{\partial \rho_{\alpha_1\alpha_2}}{\partial t}\right|_{\rm s} = \frac{1}{2} \sum_{\alpha'\alpha'_1\alpha'_2} \left(\left(\delta_{\alpha_1\alpha'} - \rho_{\alpha_1\alpha'}\right)
P^{ }_{\alpha'\alpha_2,\alpha'_1\alpha'_2} \rho_{\alpha'_1\alpha'_2} 
- 
\left(\delta_{\alpha'\alpha'_1} - \rho_{\alpha'\alpha'_1}\right) P^{*}_{\alpha'\alpha'_1,\alpha_1\alpha'_2} \rho_{\alpha'_2\alpha_2}\right)\ +\ {\rm H.c.}\ ,
\end{equation}
\end{widetext}
where, $P^{ }_{\alpha_1\alpha_2,\alpha'_1\alpha'_2}$ are generalized scattering rates, whose explicit form is given in Ref.~[\onlinecite{Rosati14e}]. Here, the nonlinearity factors $(\delta_{\alpha_1\alpha_2} - \rho_{\alpha_1\alpha_2})$ can be regarded as the quantum-mechanical generalization of the Pauli factors of the conventional Boltzmann theory [see Eq.~(\ref{BTE2})].

In order to get the desired Wigner-function version of the density-matrix scattering superoperator in (\ref{DMEsbis}), the crucial step is once again to apply to the latter the Weyl-Wigner transform  (\ref{WWT}) together with its inverse in (\ref{WWTinv}).
The resulting Wigner-function scattering superoperator his still described by the in- minus-out structure in (\ref{BTE3}), provided to replace the local terms of the semiclassical theory in (\ref{Pin}) and (\ref{Pout}) with the following nonlocal generalizations:\footnote{In spite of the strong formal similarity with the conventional Boltzmann transport theory, we stress that the generalized Wigner-function scattering rates in (\ref{Pinout}) are not necessarily positive-definite.}
\begin{eqnarray}
\lefteqn{P^{\rm in/out}(\br,\bk;\br',\bk') 
= } & &   \label{Pinout} \\
&=& \int \frac{d\br''\,d\bk''}{(2\pi)^3}\, \left(1-f(\br'',\bk'')\right)
\tilde{P}^{\rm in/out}(\br'',\bk'';\br,\bk;\br',\bk') \nonumber 
\end{eqnarray}
with
\begin{widetext} 

\begin{equation}\label{Pinbis}
\tilde{P}^{\rm in}(\br'',\bk'';\br,\bk;\br',\bk') 
= 
\frac{1}{(2\pi)^3}
\sum_{\alpha_1\alpha_2\alpha'\alpha_1'\alpha_2'} 
\Re\left\{
W^{ }_{\alpha_1\alpha_2}(\br,\bk) 
W^*_{\alpha_1\alpha'}(\br'',\bk'')  
P^{ }_{\alpha'\alpha_2,\alpha_1'\alpha_2'} 
W^*_{\alpha_1'\alpha_2'}(\br',\bk') 
\right\}
\end{equation}
and
\begin{equation}\label{Poutbis}
\tilde{P}^{{\rm out}}(\br'',\bk'';\br,\bk;\br',\bk') 
= 
\frac{1}{(2\pi)^3}
\sum_{\alpha_1\alpha_2\alpha'\alpha_1'\alpha_2'}  
\Re\left\{
W^{ }_{\alpha_1\alpha_2}(\br,\bk) 
W^*_{\alpha'\alpha_1'}(\br'',\bk'') 
P^*_{\alpha'\alpha_1',\alpha_1\alpha_2'} 
W^*_{\alpha_2'\alpha_2}(\br',\bk') 
\right\}\ .
\end{equation}
\end{widetext} 

The proposed quantum-mechanical generalization of the standard Boltzmann collision term in (\ref{BTE1}) is thus intrinsically nonlocal. In particular, comparing Eq.~(\ref{Pinout}) with its semiclassical counterpart in (\ref{BTE2}), it is evident that the action of the Pauli exclusion principle within the Wigner phase-space is itself nonlocal: the generalized in and out scattering rates for a given transition $\br,\bk \to \br',\bk'$ depend on the value of the Wigner function in any other phase-space point $\br'',\bk''$ via the Pauli factor $1-f(\br'',\bk'')$. 
A detailed investigation of such Pauli-blocking nonlocality is however outside the scope of the present work.

In the low-density limit ($f(\br,\bk) \to 0$), the proposed scattering model in (\ref{Pinout}) reduces to:
\begin{widetext} 
\begin{equation}\label{Pinter}
P^{\rm in}(\br,\bk;\br',\bk') 
= 
\frac{1}{(2\pi)^3}
\sum_{\alpha_1\alpha_2\alpha_1'\alpha_2'} 
\Re\left\{
W^{ }_{\alpha_1\alpha_2}(\br,\bk) 
P^{ }_{\alpha_1\alpha_2,\alpha_1'\alpha_2'} 
W^*_{\alpha_1'\alpha_2'}(\br',\bk') 
\right\}
\end{equation}
and
\begin{equation}\label{Poutter}
P^{{\rm out}}(\br,\bk;\br',\bk') 
= 
\frac{1}{(2\pi)^3}
\sum_{\alpha_1\alpha_2\alpha_1'\alpha_2'}  
\Re\left\{
W^{ }_{\alpha_1\alpha_2}(\br,\bk) 
P^*_{\alpha_1'\alpha_1',\alpha_1\alpha_2'} 
W^*_{\alpha_2'\alpha_2}(\br',\bk') 
\right\}\ .
\end{equation}
\end{widetext} 

\begin{figure}
\centering
\includegraphics[width=\columnwidth]{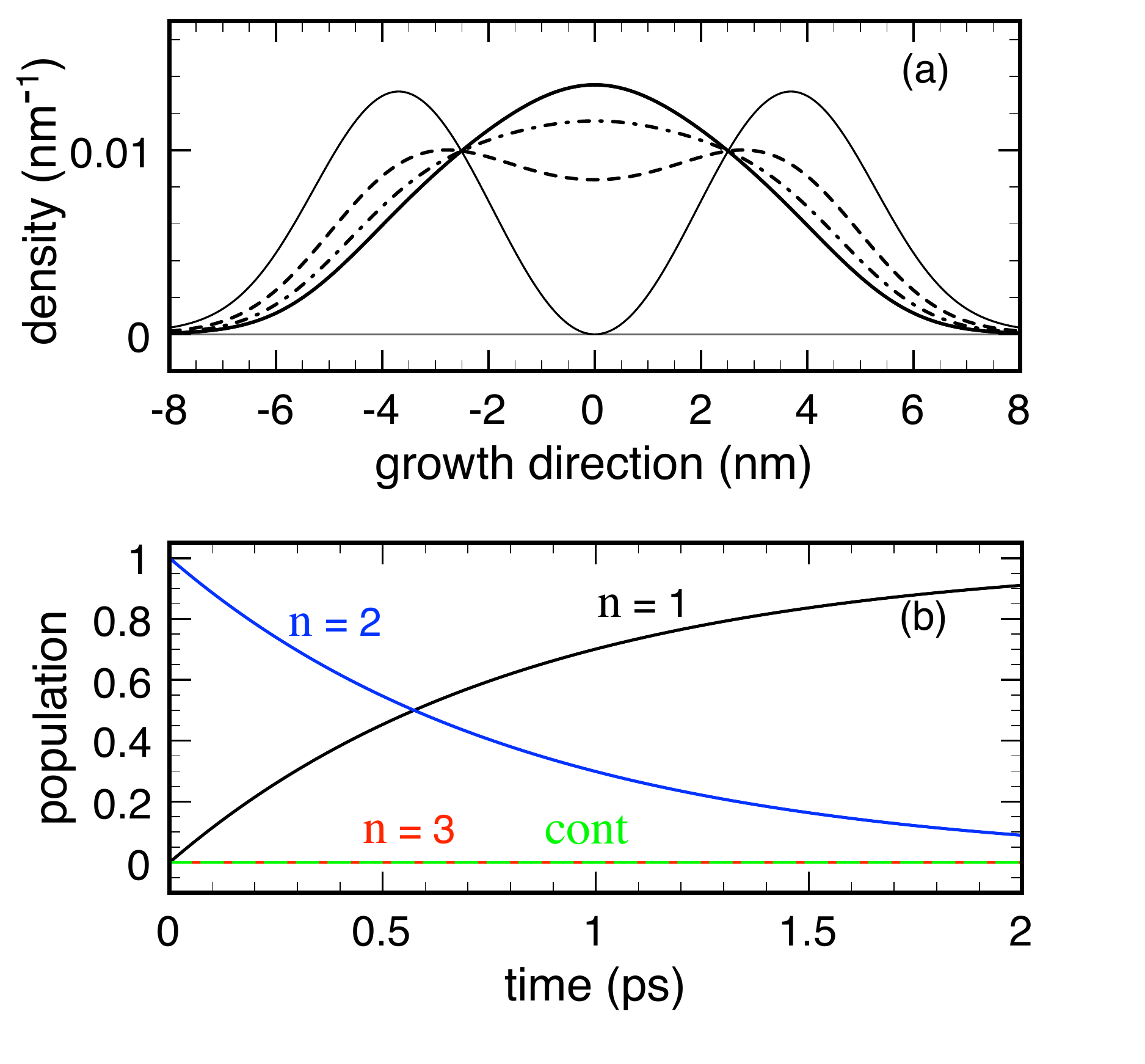}
\caption{(Color online)
Same as in Fig.~\ref{Fig4} but replacing the local Boltzmann model in (\ref{BTE1}) with the proposed nonlocal model in (\ref{Pinout}) (see text).
}
\label{Fig7}
\end{figure}

In order to test the quality of the proposed nonlocal scattering superoperator, we have repeated the simulated experiment of Fig.~\ref{Fig4} replacing the Boltzmann collision term in (\ref{BTE1}) with the nonlocal scattering model in (\ref{Pinout}); the resulting energy-dissipation scenario is shown in Fig.~\ref{Fig7}.
As one can see, opposite to the unphysical behaviors obtained via the semiclassical Boltzmann model (see Figs.~\ref{Fig4} and \ref{Fig5}), here both the spatial carrier distributions (panel a) and the corresponding subband populations (panel b) are always positive-definite; moreover, as for the case of the physically sound results of the RTA model M1 [see panels (a) in Figs.~\ref{Fig2} and \ref{Fig3}], we deal again with a correct relaxation dynamics toward the equilibrium state (solid curve in panel a) without unphysical interlevel relaxation couplings and continuum populations. Moreover, due to the trace-preserving character of the original density-matrix model in (\ref{DMEsbis}), the proposed nonlocal scattering superoperator in (\ref{Pinout}) is intrinsically charge-conserving.

\section{Summary and conclusions}\label{s-SC}

The widespread use of local scattering models, namely RTA and Boltzmann-like schemes, relies on their intuitive simplicity and easiness of implementation. In this paper we have studied their application to the Wigner-function formalism in order to characterize electronic dissipation and decoherence in semiconductor nanostructures. 

Our analysis has shown that, despite the formal similarity of the Wigner Equation to the Boltzmann transport one, when such local scattering models are applied to the Wigner function in the same way as it is done for the semiclassical Boltzmann distribution [see Eqs.(\ref{RTA}) and (\ref{BTE1})], unphysical results may be obtained; in particular, in striking contrast to the semiclassical case, one deals with anomalous suppression of intersubband relaxation, incorrect thermalization dynamics, and violation of probability-density positivity. 
We have shown that this is due (i) to the intrinsically nonlocal character of the fully quantum mechanical Wigner-function formalism, and (ii) to the bulk-like character of such semiclassical scattering models. 

Exploiting the Weyl-Wigner transform, we have then proposed a quantum-mechanical generalization both of the RTA scheme and of the Boltzmann collision term; the latter are nonlocal in space and energy [see Eqs.(\ref{RTAnew}) and (\ref{Pinout})] and guarantee positive probability densities. 

Our investigation allows us to draw the following two basic conclusions:
\begin{itemize}
\item[(i)\ ] 
Within the Wigner-function formalism the only reliable, i.e., physically correct, local scattering model is the RTA model M1 [see panels (a) in Figs.~\ref{Fig2} and \ref{Fig3}], corresponding to a constant (i.e., space- and energy-independent) relaxation rate $\Gamma_0$. In contrast, any refined version of the RTA model (based on space and/or energy dependent relaxation rates) [see panels (b) and (c) in Figs.~\ref{Fig2} and \ref{Fig3}] or any Boltzmann-like treatment (see Figs.~\ref{Fig4} and \ref{Fig5}) may lead to physically incorrect results.
\item[(ii)\ ] 
The density-matrix picture is the most natural formalism for the description of energy dissipation and decoherence; indeed, as discussed in detail in Ref.~[\onlinecite{Rosati14e}], the latter allows for a rigorous quantum-mechanical derivation of Markovian scattering superoperators [see Eq.~(\ref{DMEsbis})].
For systems with spatial open boundaries, the Wigner-function picture is generally preferable; it is however important to stress that the correct procedure is to derive the Wigner-function scattering superoperator via a Weyl-Wigner transform of its density-matrix counterpart. Any naive inclusion of semiclassical-like models may lead to the pathological behaviors previously discussed.
\end{itemize}
\par\noindent

We conclude by outlining possible future developments of our investigation. Here we have considered the case of static electric fields, which are described by a scalar potential $V$ entering the deterministic term (\ref{WTEd}) of the Wigner transport equation via the Weyl-Wigner potential in (\ref{VWW}). This situation properly describes quantum-transport phenomena in a wide class of non-magnetic semiconductor nanodevices operating in steady-state conditions; indeed, also in the presence of significant carrier concentrations, the Wigner transport equation in (\ref{WTE}) may be coupled to a corresponding Poisson equation for the scalar potential via so-called Wigner-Poisson simulation schemes.~\cite{McLennan91a} However, in the presence of magnetic fields, the description must necessarily  invoke a vector potential as well, including its possible time-dependence. In that case, the problem of gauge invariance of the Wigner function arises. It is well known\cite{Badziag85a,Serimaa86a} that a physical (i.e. gauge-independent) Wigner function $f(\br,\bk)={\rm tr}[\hat{W}(\br,\bk) \hat{\rho}]$ is obtained by modifying the Wigner operator $\hat{W}(\br,\bk)$ in Eq.(\ref{W}) via electromagnetic-potential terms that compensate for the gauge-dependence of the density matrix  $\hat{\rho}$. 
As a consequence, the deterministic term (\ref{WTEd}) of the Wigner transport equation gets modified, as described in Refs.[\onlinecite{Badziag85a}] and [\onlinecite{Serimaa86a}]. The nonlocal scattering models proposed here for the scattering term (\ref{NLSS}) of the Wigner Transport equation can thus be generalized to the presence of time-dependent electromagnetic fields, at least as long as the time-scales of the typical scattering mechanisms are shorter than the time variation of the electromagnetic fields.

We finally observe that nonlocality effects also arise in other aspects of quantum transport in nanostructures, such as the contacts with electrodes. For that problem, it has been shown\cite{Rosati13a} that the semiclassical inflow boundary condition scheme generally fails, and that the boundary values of the unknown Wigner function must be  chosen suitably to the specific device under examination;
again, the combination of a fully quantum-mechanical treatment of the system dynamics with a semiclassical injection model may produce unphysical results.
Also in this case,  the most natural strategy to overcome these limitations is to replace the semiclassical boundary-condition scheme with a density-matrix-based device-reservoir coupling model.\cite{Dolcini13a}

\begin{acknowledgments}

We are grateful to Roberto Rosati for stimulating and fruitful discussions.
Computational resources were provided by HPC@PoliTo, a project of Academic Computing of the Politecnico di Torino (www.hpc.polito.it).

\end{acknowledgments}



\begin{thebibliography}{63}%
\makeatletter
\providecommand \@ifxundefined [1]{%
 \@ifx{#1\undefined}
}%
\providecommand \@ifnum [1]{%
 \ifnum #1\expandafter \@firstoftwo
 \else \expandafter \@secondoftwo
 \fi
}%
\providecommand \@ifx [1]{%
 \ifx #1\expandafter \@firstoftwo
 \else \expandafter \@secondoftwo
 \fi
}%
\providecommand \natexlab [1]{#1}%
\providecommand \enquote  [1]{``#1''}%
\providecommand \bibnamefont  [1]{#1}%
\providecommand \bibfnamefont [1]{#1}%
\providecommand \citenamefont [1]{#1}%
\providecommand \href@noop [0]{\@secondoftwo}%
\providecommand \href [0]{\begingroup \@sanitize@url \@href}%
\providecommand \@href[1]{\@@startlink{#1}\@@href}%
\providecommand \@@href[1]{\endgroup#1\@@endlink}%
\providecommand \@sanitize@url [0]{\catcode `\\12\catcode `\$12\catcode
  `\&12\catcode `\#12\catcode `\^12\catcode `\_12\catcode `\%12\relax}%
\providecommand \@@startlink[1]{}%
\providecommand \@@endlink[0]{}%
\providecommand \url  [0]{\begingroup\@sanitize@url \@url }%
\providecommand \@url [1]{\endgroup\@href {#1}{\urlprefix }}%
\providecommand \urlprefix  [0]{URL }%
\providecommand \Eprint [0]{\href }%
\providecommand \doibase [0]{http://dx.doi.org/}%
\providecommand \selectlanguage [0]{\@gobble}%
\providecommand \bibinfo  [0]{\@secondoftwo}%
\providecommand \bibfield  [0]{\@secondoftwo}%
\providecommand \translation [1]{[#1]}%
\providecommand \BibitemOpen [0]{}%
\providecommand \bibitemStop [0]{}%
\providecommand \bibitemNoStop [0]{.\EOS\space}%
\providecommand \EOS [0]{\spacefactor3000\relax}%
\providecommand \BibitemShut  [1]{\csname bibitem#1\endcsname}%
\let\auto@bib@innerbib\@empty
\bibitem [{\citenamefont {Jacoboni}\ and\ \citenamefont
  {Lugli}(1989)}]{b-Jacoboni89}%
  \BibitemOpen
  \bibfield  {author} {\bibinfo {author} {\bibfnamefont {C.}~\bibnamefont
  {Jacoboni}}\ and\ \bibinfo {author} {\bibfnamefont {P.}~\bibnamefont
  {Lugli}},\ }\href@noop {} {\emph {\bibinfo {title} {The Monte Carlo Method
  for Semiconductor Device Simulation}}}\ (\bibinfo  {publisher} {Springer},\
  \bibinfo {year} {1989})\BibitemShut {NoStop}%
\bibitem [{\citenamefont {Bastard}(1988)}]{b-Bastard88}%
  \BibitemOpen
  \bibfield  {author} {\bibinfo {author} {\bibfnamefont {G.}~\bibnamefont
  {Bastard}},\ }\href@noop {} {\emph {\bibinfo {title} {Wave mechanics applied
  to semiconductor heterostructures}}},\ Monographies de physique\ (\bibinfo
  {publisher} {Les {\'E}ditions de Physique},\ \bibinfo {year}
  {1988})\BibitemShut {NoStop}%
\bibitem [{\citenamefont {Frensley}(1990)}]{Frensley90a}%
  \BibitemOpen
  \bibfield  {author} {\bibinfo {author} {\bibfnamefont {W.~R.}\ \bibnamefont
  {Frensley}},\ }\href {\doibase 10.1103/RevModPhys.62.745} {\bibfield
  {journal} {\bibinfo  {journal} {Rev. Mod. Phys.}\ }\textbf {\bibinfo {volume}
  {62}},\ \bibinfo {pages} {745} (\bibinfo {year} {1990})}\BibitemShut
  {NoStop}%
\bibitem [{\citenamefont {Lent}\ \emph {et~al.}(1993)\citenamefont {Lent},
  \citenamefont {Tougaw}, \citenamefont {Porod},\ and\ \citenamefont
  {Bernstein}}]{Lent93a}%
  \BibitemOpen
  \bibfield  {author} {\bibinfo {author} {\bibfnamefont {C.~S.}\ \bibnamefont
  {Lent}}, \bibinfo {author} {\bibfnamefont {P.~D.}\ \bibnamefont {Tougaw}},
  \bibinfo {author} {\bibfnamefont {W.}~\bibnamefont {Porod}}, \ and\ \bibinfo
  {author} {\bibfnamefont {G.~H.}\ \bibnamefont {Bernstein}},\ }\href@noop {}
  {\bibfield  {journal} {\bibinfo  {journal} {Nanotechnology}\ }\textbf
  {\bibinfo {volume} {4}},\ \bibinfo {pages} {49} (\bibinfo {year}
  {1993})}\BibitemShut {NoStop}%
\bibitem [{\citenamefont {Di~Carlo}\ \emph {et~al.}(1994)\citenamefont
  {Di~Carlo}, \citenamefont {Vogl},\ and\ \citenamefont {P\"otz}}]{DiCarlo94a}%
  \BibitemOpen
  \bibfield  {author} {\bibinfo {author} {\bibfnamefont {A.}~\bibnamefont
  {Di~Carlo}}, \bibinfo {author} {\bibfnamefont {P.}~\bibnamefont {Vogl}}, \
  and\ \bibinfo {author} {\bibfnamefont {W.}~\bibnamefont {P\"otz}},\ }\href
  {\doibase 10.1103/PhysRevB.50.8358} {\bibfield  {journal} {\bibinfo
  {journal} {Phys. Rev. B}\ }\textbf {\bibinfo {volume} {50}},\ \bibinfo
  {pages} {8358} (\bibinfo {year} {1994})}\BibitemShut {NoStop}%
\bibitem [{\citenamefont {Savasta}\ and\ \citenamefont
  {Girlanda}(1996)}]{Savasta96a}%
  \BibitemOpen
  \bibfield  {author} {\bibinfo {author} {\bibfnamefont {S.}~\bibnamefont
  {Savasta}}\ and\ \bibinfo {author} {\bibfnamefont {R.}~\bibnamefont
  {Girlanda}},\ }\href {\doibase 10.1103/PhysRevLett.77.4736} {\bibfield
  {journal} {\bibinfo  {journal} {Phys. Rev. Lett.}\ }\textbf {\bibinfo
  {volume} {77}},\ \bibinfo {pages} {4736} (\bibinfo {year}
  {1996})}\BibitemShut {NoStop}%
\bibitem [{\citenamefont {Fischetti}(1999)}]{Fischetti99a}%
  \BibitemOpen
  \bibfield  {author} {\bibinfo {author} {\bibfnamefont {M.~V.}\ \bibnamefont
  {Fischetti}},\ }\href {\doibase 10.1103/PhysRevB.59.4901} {\bibfield
  {journal} {\bibinfo  {journal} {Phys. Rev. B}\ }\textbf {\bibinfo {volume}
  {59}},\ \bibinfo {pages} {4901} (\bibinfo {year} {1999})}\BibitemShut
  {NoStop}%
\bibitem [{\citenamefont {Datta}(2000)}]{Datta00a}%
  \BibitemOpen
  \bibfield  {author} {\bibinfo {author} {\bibfnamefont {S.}~\bibnamefont
  {Datta}},\ }\href {\doibase 10.1006/spmi.2000.0920} {\bibfield  {journal}
  {\bibinfo  {journal} {Superlattice. Microst.}\ }\textbf {\bibinfo {volume}
  {28}},\ \bibinfo {pages} {253} (\bibinfo {year} {2000})}\BibitemShut
  {NoStop}%
\bibitem [{\citenamefont {Rossi}\ and\ \citenamefont {Kuhn}(2002)}]{Rossi02b}%
  \BibitemOpen
  \bibfield  {author} {\bibinfo {author} {\bibfnamefont {F.}~\bibnamefont
  {Rossi}}\ and\ \bibinfo {author} {\bibfnamefont {T.}~\bibnamefont {Kuhn}},\
  }\href {\doibase 10.1103/RevModPhys.74.895} {\bibfield  {journal} {\bibinfo
  {journal} {Rev. Mod. Phys.}\ }\textbf {\bibinfo {volume} {74}},\ \bibinfo
  {pages} {895} (\bibinfo {year} {2002})}\BibitemShut {NoStop}%
\bibitem [{\citenamefont {Axt}\ and\ \citenamefont {Kuhn}(2004)}]{Axt04a}%
  \BibitemOpen
  \bibfield  {author} {\bibinfo {author} {\bibfnamefont {V.~M.}\ \bibnamefont
  {Axt}}\ and\ \bibinfo {author} {\bibfnamefont {T.}~\bibnamefont {Kuhn}},\
  }\href {\doibase 10.1088/0034-4885/67/4/R01} {\bibfield  {journal} {\bibinfo
  {journal} {Rep. Prog. Phys.}\ }\textbf {\bibinfo {volume} {67}},\ \bibinfo
  {pages} {433} (\bibinfo {year} {2004})}\BibitemShut {NoStop}%
\bibitem [{\citenamefont {Jacoboni}\ and\ \citenamefont
  {Bordone}(2004)}]{Jacoboni04a}%
  \BibitemOpen
  \bibfield  {author} {\bibinfo {author} {\bibfnamefont {C.}~\bibnamefont
  {Jacoboni}}\ and\ \bibinfo {author} {\bibfnamefont {P.}~\bibnamefont
  {Bordone}},\ }\href@noop {} {\bibfield  {journal} {\bibinfo  {journal} {Rep.
  Prog. Phys.}\ }\textbf {\bibinfo {volume} {67}},\ \bibinfo {pages} {1033}
  (\bibinfo {year} {2004})}\BibitemShut {NoStop}%
\bibitem [{\citenamefont {Pecchia}\ and\ \citenamefont
  {Di~Carlo}(2004)}]{Pecchia04a}%
  \BibitemOpen
  \bibfield  {author} {\bibinfo {author} {\bibfnamefont {A.}~\bibnamefont
  {Pecchia}}\ and\ \bibinfo {author} {\bibfnamefont {A.}~\bibnamefont
  {Di~Carlo}},\ }\href {\doibase 10.1088/0034-4885/67/8/R04} {\bibfield
  {journal} {\bibinfo  {journal} {Rep. Prog. Phys.}\ }\textbf {\bibinfo
  {volume} {67}},\ \bibinfo {pages} {1497} (\bibinfo {year}
  {2004})}\BibitemShut {NoStop}%
\bibitem [{\citenamefont {Iotti}\ and\ \citenamefont {Rossi}(2005)}]{Iotti05b}%
  \BibitemOpen
  \bibfield  {author} {\bibinfo {author} {\bibfnamefont {R.~C.}\ \bibnamefont
  {Iotti}}\ and\ \bibinfo {author} {\bibfnamefont {F.}~\bibnamefont {Rossi}},\
  }\href {\doibase 10.1088/0034-4885/68/11/R02} {\bibfield  {journal} {\bibinfo
   {journal} {Rep. Prog. Phys.}\ }\textbf {\bibinfo {volume} {68}},\ \bibinfo
  {pages} {2533} (\bibinfo {year} {2005})}\BibitemShut {NoStop}%
\bibitem [{\citenamefont {Datta}(2005)}]{b-Datta05}%
  \BibitemOpen
  \bibfield  {author} {\bibinfo {author} {\bibfnamefont {S.}~\bibnamefont
  {Datta}},\ }\href@noop {} {\emph {\bibinfo {title} {Quantum Transport: Atom
  to Transistor}}}\ (\bibinfo  {publisher} {Cambridge University Press},\
  \bibinfo {year} {2005})\BibitemShut {NoStop}%
\bibitem [{\citenamefont {Haug}\ and\ \citenamefont {Jauho}(2007)}]{b-Haug07}%
  \BibitemOpen
  \bibfield  {author} {\bibinfo {author} {\bibfnamefont {H.}~\bibnamefont
  {Haug}}\ and\ \bibinfo {author} {\bibfnamefont {A.}~\bibnamefont {Jauho}},\
  }\href@noop {} {\emph {\bibinfo {title} {Quantum Kinetics in Transport and
  Optics of Semiconductors}}}\ (\bibinfo  {publisher} {Springer},\ \bibinfo
  {year} {2007})\BibitemShut {NoStop}%
\bibitem [{\citenamefont {Jacoboni}(2010)}]{b-Jacoboni10}%
  \BibitemOpen
  \bibfield  {author} {\bibinfo {author} {\bibfnamefont {C.}~\bibnamefont
  {Jacoboni}},\ }\href@noop {} {\emph {\bibinfo {title} {Theory of Electron
  Transport in Semiconductors: A Pathway from Elementary Physics to
  Nonequilibrium Green Functions}}}\ (\bibinfo  {publisher} {Springer},\
  \bibinfo {year} {2010})\BibitemShut {NoStop}%
\bibitem [{\citenamefont {Haug}\ and\ \citenamefont {Koch}(2004)}]{b-Haug04}%
  \BibitemOpen
  \bibfield  {author} {\bibinfo {author} {\bibfnamefont {H.}~\bibnamefont
  {Haug}}\ and\ \bibinfo {author} {\bibfnamefont {S.}~\bibnamefont {Koch}},\
  }\href@noop {} {\emph {\bibinfo {title} {Quantum Theory of the Optical and
  Electronic Properties of Semiconductors}}}\ (\bibinfo  {publisher} {World
  Scientific},\ \bibinfo {year} {2004})\BibitemShut {NoStop}%
\bibitem [{\citenamefont {Rossi}(2011)}]{b-Rossi11}%
  \BibitemOpen
  \bibfield  {author} {\bibinfo {author} {\bibfnamefont {F.}~\bibnamefont
  {Rossi}},\ }\href@noop {} {\emph {\bibinfo {title} {Theory of Semiconductor
  Quantum Devices: Microscopic Modeling and Simulation Strategies}}}\ (\bibinfo
   {publisher} {Springer},\ \bibinfo {year} {2011})\BibitemShut {NoStop}%
\bibitem [{\citenamefont {Buot}(2009)}]{b-Buot09}%
  \BibitemOpen
  \bibfield  {author} {\bibinfo {author} {\bibfnamefont {F.}~\bibnamefont
  {Buot}},\ }\href@noop {} {\emph {\bibinfo {title} {Nonequilibrium quantum
  transport physics in nanosystems: foundation of computational nonequilibrium
  physics in nanoscience and nanotechnology}}}\ (\bibinfo  {publisher} {World
  Scientific},\ \bibinfo {year} {2009})\BibitemShut {NoStop}%
\bibitem [{\citenamefont {Rosati}\ \emph {et~al.}(2013)\citenamefont {Rosati},
  \citenamefont {Dolcini}, \citenamefont {Iotti},\ and\ \citenamefont
  {Rossi}}]{Rosati13a}%
  \BibitemOpen
  \bibfield  {author} {\bibinfo {author} {\bibfnamefont {R.}~\bibnamefont
  {Rosati}}, \bibinfo {author} {\bibfnamefont {F.}~\bibnamefont {Dolcini}},
  \bibinfo {author} {\bibfnamefont {R.~C.}\ \bibnamefont {Iotti}}, \ and\
  \bibinfo {author} {\bibfnamefont {F.}~\bibnamefont {Rossi}},\ }\href
  {\doibase 10.1103/PhysRevB.88.035401} {\bibfield  {journal} {\bibinfo
  {journal} {Phys. Rev. B}\ }\textbf {\bibinfo {volume} {88}},\ \bibinfo
  {pages} {035401} (\bibinfo {year} {2013})}\BibitemShut {NoStop}%
\bibitem [{\citenamefont {Taj}\ \emph {et~al.}(2009)\citenamefont {Taj},
  \citenamefont {Iotti},\ and\ \citenamefont {Rossi}}]{Taj09b}%
  \BibitemOpen
  \bibfield  {author} {\bibinfo {author} {\bibfnamefont {D.}~\bibnamefont
  {Taj}}, \bibinfo {author} {\bibfnamefont {R.~C.}\ \bibnamefont {Iotti}}, \
  and\ \bibinfo {author} {\bibfnamefont {F.}~\bibnamefont {Rossi}},\ }\href
  {\doibase 10.1140/epjb/e2009-00363-4} {\bibfield  {journal} {\bibinfo
  {journal} {Eur. Phys. J. B}\ }\textbf {\bibinfo {volume} {72}},\ \bibinfo
  {pages} {305} (\bibinfo {year} {2009})}\BibitemShut {NoStop}%
\bibitem [{\citenamefont {Rosati}\ \emph {et~al.}(2014)\citenamefont {Rosati},
  \citenamefont {Iotti}, \citenamefont {Dolcini},\ and\ \citenamefont
  {Rossi}}]{Rosati14e}%
  \BibitemOpen
  \bibfield  {author} {\bibinfo {author} {\bibfnamefont {R.}~\bibnamefont
  {Rosati}}, \bibinfo {author} {\bibfnamefont {R.~C.}\ \bibnamefont {Iotti}},
  \bibinfo {author} {\bibfnamefont {F.}~\bibnamefont {Dolcini}}, \ and\
  \bibinfo {author} {\bibfnamefont {F.}~\bibnamefont {Rossi}},\ }\href
  {\doibase 10.1103/PhysRevB.90.125140} {\bibfield  {journal} {\bibinfo
  {journal} {Phys. Rev. B}\ }\textbf {\bibinfo {volume} {90}},\ \bibinfo
  {pages} {125140} (\bibinfo {year} {2014})}\BibitemShut {NoStop}%
\bibitem [{\citenamefont {Frensley}(1986)}]{Frensley86a}%
  \BibitemOpen
  \bibfield  {author} {\bibinfo {author} {\bibfnamefont {W.~R.}\ \bibnamefont
  {Frensley}},\ }\href {\doibase 10.1103/PhysRevLett.57.2853} {\bibfield
  {journal} {\bibinfo  {journal} {Phys. Rev. Lett.}\ }\textbf {\bibinfo
  {volume} {57}},\ \bibinfo {pages} {2853} (\bibinfo {year}
  {1986})}\BibitemShut {NoStop}%
\bibitem [{\citenamefont {Kluksdahl}\ \emph {et~al.}(1989)\citenamefont
  {Kluksdahl}, \citenamefont {Kriman}, \citenamefont {Ferry},\ and\
  \citenamefont {Ringhofer}}]{Kluksdahl89a}%
  \BibitemOpen
  \bibfield  {author} {\bibinfo {author} {\bibfnamefont {N.~C.}\ \bibnamefont
  {Kluksdahl}}, \bibinfo {author} {\bibfnamefont {A.~M.}\ \bibnamefont
  {Kriman}}, \bibinfo {author} {\bibfnamefont {D.~K.}\ \bibnamefont {Ferry}}, \
  and\ \bibinfo {author} {\bibfnamefont {C.}~\bibnamefont {Ringhofer}},\ }\href
  {\doibase 10.1103/PhysRevB.39.7720} {\bibfield  {journal} {\bibinfo
  {journal} {Phys. Rev. B}\ }\textbf {\bibinfo {volume} {39}},\ \bibinfo
  {pages} {7720} (\bibinfo {year} {1989})}\BibitemShut {NoStop}%
\bibitem [{\citenamefont {Buot}\ and\ \citenamefont {Jensen}(1990)}]{Buot90a}%
  \BibitemOpen
  \bibfield  {author} {\bibinfo {author} {\bibfnamefont {F.~A.}\ \bibnamefont
  {Buot}}\ and\ \bibinfo {author} {\bibfnamefont {K.~L.}\ \bibnamefont
  {Jensen}},\ }\href {\doibase 10.1103/PhysRevB.42.9429} {\bibfield  {journal}
  {\bibinfo  {journal} {Phys. Rev. B}\ }\textbf {\bibinfo {volume} {42}},\
  \bibinfo {pages} {9429} (\bibinfo {year} {1990})}\BibitemShut {NoStop}%
\bibitem [{\citenamefont {Jensen}\ and\ \citenamefont
  {Buot}(1990)}]{Jensen90a}%
  \BibitemOpen
  \bibfield  {author} {\bibinfo {author} {\bibfnamefont {K.}~\bibnamefont
  {Jensen}}\ and\ \bibinfo {author} {\bibfnamefont {F.}~\bibnamefont {Buot}},\
  }\href {\doibase 10.1063/1.345828} {\bibfield  {journal} {\bibinfo  {journal}
  {J. Appl. Phys.}\ }\textbf {\bibinfo {volume} {67}},\ \bibinfo {pages} {7602}
  (\bibinfo {year} {1990})}\BibitemShut {NoStop}%
\bibitem [{\citenamefont {Miller}\ and\ \citenamefont
  {Neikirk}(1991)}]{Miller91a}%
  \BibitemOpen
  \bibfield  {author} {\bibinfo {author} {\bibfnamefont {D.~R.}\ \bibnamefont
  {Miller}}\ and\ \bibinfo {author} {\bibfnamefont {D.~P.}\ \bibnamefont
  {Neikirk}},\ }\href {\doibase http://dx.doi.org/10.1063/1.104741} {\bibfield
  {journal} {\bibinfo  {journal} {Appl. Phys. Lett.}\ }\textbf {\bibinfo
  {volume} {58}},\ \bibinfo {pages} {2803} (\bibinfo {year}
  {1991})}\BibitemShut {NoStop}%
\bibitem [{\citenamefont {McLennan}\ \emph {et~al.}(1991)\citenamefont
  {McLennan}, \citenamefont {Lee},\ and\ \citenamefont {Datta}}]{McLennan91a}%
  \BibitemOpen
  \bibfield  {author} {\bibinfo {author} {\bibfnamefont {M.~J.}\ \bibnamefont
  {McLennan}}, \bibinfo {author} {\bibfnamefont {Y.}~\bibnamefont {Lee}}, \
  and\ \bibinfo {author} {\bibfnamefont {S.}~\bibnamefont {Datta}},\ }\href
  {\doibase 10.1103/PhysRevB.43.13846} {\bibfield  {journal} {\bibinfo
  {journal} {Phys. Rev. B}\ }\textbf {\bibinfo {volume} {43}},\ \bibinfo
  {pages} {13846} (\bibinfo {year} {1991})}\BibitemShut {NoStop}%
\bibitem [{\citenamefont {Tso}\ and\ \citenamefont {Horing}(1991)}]{Tso91a}%
  \BibitemOpen
  \bibfield  {author} {\bibinfo {author} {\bibfnamefont {H.~C.}\ \bibnamefont
  {Tso}}\ and\ \bibinfo {author} {\bibfnamefont {N.~J.~M.}\ \bibnamefont
  {Horing}},\ }\href {\doibase 10.1103/PhysRevB.44.11358} {\bibfield  {journal}
  {\bibinfo  {journal} {Phys. Rev. B}\ }\textbf {\bibinfo {volume} {44}},\
  \bibinfo {pages} {11358} (\bibinfo {year} {1991})}\BibitemShut {NoStop}%
\bibitem [{\citenamefont {Gullapalli}\ \emph {et~al.}(1994)\citenamefont
  {Gullapalli}, \citenamefont {Miller},\ and\ \citenamefont
  {Neikirk}}]{Gullapalli94a}%
  \BibitemOpen
  \bibfield  {author} {\bibinfo {author} {\bibfnamefont {K.~K.}\ \bibnamefont
  {Gullapalli}}, \bibinfo {author} {\bibfnamefont {D.~R.}\ \bibnamefont
  {Miller}}, \ and\ \bibinfo {author} {\bibfnamefont {D.~P.}\ \bibnamefont
  {Neikirk}},\ }\href {\doibase 10.1103/PhysRevB.49.2622} {\bibfield  {journal}
  {\bibinfo  {journal} {Phys. Rev. B}\ }\textbf {\bibinfo {volume} {49}},\
  \bibinfo {pages} {2622} (\bibinfo {year} {1994})}\BibitemShut {NoStop}%
\bibitem [{\citenamefont {Fernando}\ and\ \citenamefont
  {Frensley}(1995)}]{Fernando95a}%
  \BibitemOpen
  \bibfield  {author} {\bibinfo {author} {\bibfnamefont {C.~L.}\ \bibnamefont
  {Fernando}}\ and\ \bibinfo {author} {\bibfnamefont {W.~R.}\ \bibnamefont
  {Frensley}},\ }\href {\doibase 10.1103/PhysRevB.52.5092} {\bibfield
  {journal} {\bibinfo  {journal} {Phys. Rev. B}\ }\textbf {\bibinfo {volume}
  {52}},\ \bibinfo {pages} {5092} (\bibinfo {year} {1995})}\BibitemShut
  {NoStop}%
\bibitem [{\citenamefont {El~Sayed}\ \emph {et~al.}(1998)\citenamefont
  {El~Sayed}, \citenamefont {Kenrow},\ and\ \citenamefont
  {Stanton}}]{ElSayed98a}%
  \BibitemOpen
  \bibfield  {author} {\bibinfo {author} {\bibfnamefont {K.}~\bibnamefont
  {El~Sayed}}, \bibinfo {author} {\bibfnamefont {J.~A.}\ \bibnamefont
  {Kenrow}}, \ and\ \bibinfo {author} {\bibfnamefont {C.~J.}\ \bibnamefont
  {Stanton}},\ }\href {\doibase 10.1103/PhysRevB.57.12369} {\bibfield
  {journal} {\bibinfo  {journal} {Phys. Rev. B}\ }\textbf {\bibinfo {volume}
  {57}},\ \bibinfo {pages} {12369} (\bibinfo {year} {1998})}\BibitemShut
  {NoStop}%
\bibitem [{\citenamefont {Kim}\ and\ \citenamefont {Lee}(2001)}]{Kim01a}%
  \BibitemOpen
  \bibfield  {author} {\bibinfo {author} {\bibfnamefont {K.-Y.}\ \bibnamefont
  {Kim}}\ and\ \bibinfo {author} {\bibfnamefont {B.}~\bibnamefont {Lee}},\
  }\href {\doibase 10.1103/PhysRevB.64.115304} {\bibfield  {journal} {\bibinfo
  {journal} {Phys. Rev. B}\ }\textbf {\bibinfo {volume} {64}},\ \bibinfo
  {pages} {115304} (\bibinfo {year} {2001})}\BibitemShut {NoStop}%
\bibitem [{\citenamefont {Pascoli}\ \emph {et~al.}(1998)\citenamefont
  {Pascoli}, \citenamefont {Bordone}, \citenamefont {Brunetti},\ and\
  \citenamefont {Jacoboni}}]{Pascoli98a}%
  \BibitemOpen
  \bibfield  {author} {\bibinfo {author} {\bibfnamefont {M.}~\bibnamefont
  {Pascoli}}, \bibinfo {author} {\bibfnamefont {P.}~\bibnamefont {Bordone}},
  \bibinfo {author} {\bibfnamefont {R.}~\bibnamefont {Brunetti}}, \ and\
  \bibinfo {author} {\bibfnamefont {C.}~\bibnamefont {Jacoboni}},\ }\href
  {\doibase 10.1103/PhysRevB.58.3503} {\bibfield  {journal} {\bibinfo
  {journal} {Phys. Rev. B}\ }\textbf {\bibinfo {volume} {58}},\ \bibinfo
  {pages} {3503} (\bibinfo {year} {1998})}\BibitemShut {NoStop}%
\bibitem [{\citenamefont {Nedjalkov}\ \emph {et~al.}(2004)\citenamefont
  {Nedjalkov}, \citenamefont {Kosina}, \citenamefont {Selberherr},
  \citenamefont {Ringhofer},\ and\ \citenamefont {Ferry}}]{Nedjalkov04a}%
  \BibitemOpen
  \bibfield  {author} {\bibinfo {author} {\bibfnamefont {M.}~\bibnamefont
  {Nedjalkov}}, \bibinfo {author} {\bibfnamefont {H.}~\bibnamefont {Kosina}},
  \bibinfo {author} {\bibfnamefont {S.}~\bibnamefont {Selberherr}}, \bibinfo
  {author} {\bibfnamefont {C.}~\bibnamefont {Ringhofer}}, \ and\ \bibinfo
  {author} {\bibfnamefont {D.~K.}\ \bibnamefont {Ferry}},\ }\href {\doibase
  10.1103/PhysRevB.70.115319} {\bibfield  {journal} {\bibinfo  {journal} {Phys.
  Rev. B}\ }\textbf {\bibinfo {volume} {70}},\ \bibinfo {pages} {115319}
  (\bibinfo {year} {2004})}\BibitemShut {NoStop}%
\bibitem [{\citenamefont {Nedjalkov}\ \emph {et~al.}(2006)\citenamefont
  {Nedjalkov}, \citenamefont {Vasileska}, \citenamefont {Ferry}, \citenamefont
  {Jacoboni}, \citenamefont {Ringhofer}, \citenamefont {Dimov},\ and\
  \citenamefont {Palankovski}}]{Nedjalkov06a}%
  \BibitemOpen
  \bibfield  {author} {\bibinfo {author} {\bibfnamefont {M.}~\bibnamefont
  {Nedjalkov}}, \bibinfo {author} {\bibfnamefont {D.}~\bibnamefont
  {Vasileska}}, \bibinfo {author} {\bibfnamefont {D.~K.}\ \bibnamefont
  {Ferry}}, \bibinfo {author} {\bibfnamefont {C.}~\bibnamefont {Jacoboni}},
  \bibinfo {author} {\bibfnamefont {C.}~\bibnamefont {Ringhofer}}, \bibinfo
  {author} {\bibfnamefont {I.}~\bibnamefont {Dimov}}, \ and\ \bibinfo {author}
  {\bibfnamefont {V.}~\bibnamefont {Palankovski}},\ }\href {\doibase
  10.1103/PhysRevB.74.035311} {\bibfield  {journal} {\bibinfo  {journal} {Phys.
  Rev. B}\ }\textbf {\bibinfo {volume} {74}},\ \bibinfo {pages} {035311}
  (\bibinfo {year} {2006})}\BibitemShut {NoStop}%
\bibitem [{\citenamefont {Weetman}\ and\ \citenamefont
  {Wartak}(2007)}]{Weetman07a}%
  \BibitemOpen
  \bibfield  {author} {\bibinfo {author} {\bibfnamefont {P.}~\bibnamefont
  {Weetman}}\ and\ \bibinfo {author} {\bibfnamefont {M.~S.}\ \bibnamefont
  {Wartak}},\ }\href {\doibase 10.1103/PhysRevB.76.035332} {\bibfield
  {journal} {\bibinfo  {journal} {Phys. Rev. B}\ }\textbf {\bibinfo {volume}
  {76}},\ \bibinfo {pages} {035332} (\bibinfo {year} {2007})}\BibitemShut
  {NoStop}%
\bibitem [{\citenamefont {Querlioz}\ \emph {et~al.}(2008)\citenamefont
  {Querlioz}, \citenamefont {Saint-Martin}, \citenamefont {Bournel},\ and\
  \citenamefont {Dollfus}}]{Querlioz08a}%
  \BibitemOpen
  \bibfield  {author} {\bibinfo {author} {\bibfnamefont {D.}~\bibnamefont
  {Querlioz}}, \bibinfo {author} {\bibfnamefont {J.}~\bibnamefont
  {Saint-Martin}}, \bibinfo {author} {\bibfnamefont {A.}~\bibnamefont
  {Bournel}}, \ and\ \bibinfo {author} {\bibfnamefont {P.}~\bibnamefont
  {Dollfus}},\ }\href {\doibase 10.1103/PhysRevB.78.165306} {\bibfield
  {journal} {\bibinfo  {journal} {Phys. Rev. B}\ }\textbf {\bibinfo {volume}
  {78}},\ \bibinfo {pages} {165306} (\bibinfo {year} {2008})}\BibitemShut
  {NoStop}%
\bibitem [{\citenamefont {Morandi}(2009)}]{Morandi09a}%
  \BibitemOpen
  \bibfield  {author} {\bibinfo {author} {\bibfnamefont {O.}~\bibnamefont
  {Morandi}},\ }\href {\doibase 10.1103/PhysRevB.80.024301} {\bibfield
  {journal} {\bibinfo  {journal} {Phys. Rev. B}\ }\textbf {\bibinfo {volume}
  {80}},\ \bibinfo {pages} {024301} (\bibinfo {year} {2009})}\BibitemShut
  {NoStop}%
\bibitem [{\citenamefont {W\'{o}jcik}\ \emph {et~al.}(2009)\citenamefont
  {W\'{o}jcik}, \citenamefont {Spisak}, \citenamefont {Wo{\l}oszyn},\ and\
  \citenamefont {Adamowski}}]{Wojcik09a}%
  \BibitemOpen
  \bibfield  {author} {\bibinfo {author} {\bibfnamefont {P.}~\bibnamefont
  {W\'{o}jcik}}, \bibinfo {author} {\bibfnamefont {B.}~\bibnamefont {Spisak}},
  \bibinfo {author} {\bibfnamefont {M.}~\bibnamefont {Wo{\l}oszyn}}, \ and\
  \bibinfo {author} {\bibfnamefont {J.}~\bibnamefont {Adamowski}},\ }\href
  {\doibase 10.1088/0268-1242/24/9/095012} {\bibfield  {journal} {\bibinfo
  {journal} {Semicond. Sci. Tech.}\ }\textbf {\bibinfo {volume} {24}},\
  \bibinfo {pages} {095012} (\bibinfo {year} {2009})}\BibitemShut {NoStop}%
\bibitem [{\citenamefont {Barraud}(2009)}]{Barraud09a}%
  \BibitemOpen
  \bibfield  {author} {\bibinfo {author} {\bibfnamefont {S.}~\bibnamefont
  {Barraud}},\ }\href {\doibase 10.1063/1.3226856} {\bibfield  {journal}
  {\bibinfo  {journal} {J. Appl. Phys.}\ }\textbf {\bibinfo {volume} {106}},\
  \bibinfo {pages} {063714} (\bibinfo {year} {2009})}\BibitemShut {NoStop}%
\bibitem [{\citenamefont {Yoder}\ \emph {et~al.}(2010)\citenamefont {Yoder},
  \citenamefont {Grupen},\ and\ \citenamefont {Smith}}]{Yoder10a}%
  \BibitemOpen
  \bibfield  {author} {\bibinfo {author} {\bibfnamefont {P.~D.}\ \bibnamefont
  {Yoder}}, \bibinfo {author} {\bibfnamefont {M.}~\bibnamefont {Grupen}}, \
  and\ \bibinfo {author} {\bibfnamefont {R.}~\bibnamefont {Smith}},\ }\href
  {\doibase 10.1109/TED.2010.2081672} {\bibfield  {journal} {\bibinfo
  {journal} {IEEE Trans. Electron Devices}\ }\textbf {\bibinfo {volume} {57}},\
  \bibinfo {pages} {3265} (\bibinfo {year} {2010})}\BibitemShut {NoStop}%
\bibitem [{\citenamefont {\'Alvaro}\ and\ \citenamefont
  {Bonilla}(2010)}]{Alvaro10a}%
  \BibitemOpen
  \bibfield  {author} {\bibinfo {author} {\bibfnamefont {M.}~\bibnamefont
  {\'Alvaro}}\ and\ \bibinfo {author} {\bibfnamefont {L.~L.}\ \bibnamefont
  {Bonilla}},\ }\href {\doibase 10.1103/PhysRevB.82.035305} {\bibfield
  {journal} {\bibinfo  {journal} {Phys. Rev. B}\ }\textbf {\bibinfo {volume}
  {82}},\ \bibinfo {pages} {035305} (\bibinfo {year} {2010})}\BibitemShut
  {NoStop}%
\bibitem [{\citenamefont {Savio}\ and\ \citenamefont
  {Poncet}(2011)}]{Savio11a}%
  \BibitemOpen
  \bibfield  {author} {\bibinfo {author} {\bibfnamefont {A.}~\bibnamefont
  {Savio}}\ and\ \bibinfo {author} {\bibfnamefont {A.}~\bibnamefont {Poncet}},\
  }\href {\doibase 10.1063/1.3526969} {\bibfield  {journal} {\bibinfo
  {journal} {J. Appl. Phys.}\ }\textbf {\bibinfo {volume} {109}},\ \bibinfo
  {pages} {033713} (\bibinfo {year} {2011})}\BibitemShut {NoStop}%
\bibitem [{\citenamefont {Trovato}\ and\ \citenamefont
  {Reggiani}(2011)}]{Trovato11a}%
  \BibitemOpen
  \bibfield  {author} {\bibinfo {author} {\bibfnamefont {M.}~\bibnamefont
  {Trovato}}\ and\ \bibinfo {author} {\bibfnamefont {L.}~\bibnamefont
  {Reggiani}},\ }\href {\doibase 10.1103/PhysRevE.84.061147} {\bibfield
  {journal} {\bibinfo  {journal} {Phys. Rev. E}\ }\textbf {\bibinfo {volume}
  {84}},\ \bibinfo {pages} {061147} (\bibinfo {year} {2011})}\BibitemShut
  {NoStop}%
\bibitem [{\citenamefont {Sellier}\ \emph {et~al.}(2014)\citenamefont
  {Sellier}, \citenamefont {Amoroso}, \citenamefont {Nedjalkov}, \citenamefont
  {Selberherr}, \citenamefont {Asenov},\ and\ \citenamefont
  {Dimov}}]{Sellier14a}%
  \BibitemOpen
  \bibfield  {author} {\bibinfo {author} {\bibfnamefont {J.}~\bibnamefont
  {Sellier}}, \bibinfo {author} {\bibfnamefont {S.}~\bibnamefont {Amoroso}},
  \bibinfo {author} {\bibfnamefont {M.}~\bibnamefont {Nedjalkov}}, \bibinfo
  {author} {\bibfnamefont {S.}~\bibnamefont {Selberherr}}, \bibinfo {author}
  {\bibfnamefont {A.}~\bibnamefont {Asenov}}, \ and\ \bibinfo {author}
  {\bibfnamefont {I.}~\bibnamefont {Dimov}},\ }\href {\doibase
  http://dx.doi.org/10.1016/j.physa.2013.12.045} {\bibfield  {journal}
  {\bibinfo  {journal} {Physica A}\ }\textbf {\bibinfo {volume} {398}},\
  \bibinfo {pages} {194 } (\bibinfo {year} {2014})}\BibitemShut {NoStop}%
\bibitem [{\citenamefont {Sellier}\ and\ \citenamefont
  {Dimov}(2014)}]{Sellier14b}%
  \BibitemOpen
  \bibfield  {author} {\bibinfo {author} {\bibfnamefont {J.}~\bibnamefont
  {Sellier}}\ and\ \bibinfo {author} {\bibfnamefont {I.}~\bibnamefont
  {Dimov}},\ }\href {\doibase http://dx.doi.org/10.1016/j.physa.2014.03.065}
  {\bibfield  {journal} {\bibinfo  {journal} {Physica A}\ }\textbf {\bibinfo
  {volume} {406}},\ \bibinfo {pages} {185 } (\bibinfo {year}
  {2014})}\BibitemShut {NoStop}%
\bibitem [{\citenamefont {Jonasson}\ and\ \citenamefont
  {Knezevic}(2015)}]{Jonasson15a}%
  \BibitemOpen
  \bibfield  {author} {\bibinfo {author} {\bibfnamefont {O.}~\bibnamefont
  {Jonasson}}\ and\ \bibinfo {author} {\bibfnamefont {I.}~\bibnamefont
  {Knezevic}},\ }\href {\doibase 10.1007/s10825-015-0734-9} {\bibfield
  {journal} {\bibinfo  {journal} {J. Comput. Electron.}\ }\textbf {\bibinfo
  {volume} {14}},\ \bibinfo {pages} {879} (\bibinfo {year} {2015})}\BibitemShut
  {NoStop}%
\bibitem [{\citenamefont {Hamerly}\ and\ \citenamefont
  {Mabuchi}(2015)}]{Hamerly15a}%
  \BibitemOpen
  \bibfield  {author} {\bibinfo {author} {\bibfnamefont {R.}~\bibnamefont
  {Hamerly}}\ and\ \bibinfo {author} {\bibfnamefont {H.}~\bibnamefont
  {Mabuchi}},\ }\href {\doibase 10.1103/PhysRevA.92.023819} {\bibfield
  {journal} {\bibinfo  {journal} {Phys. Rev. A}\ }\textbf {\bibinfo {volume}
  {92}},\ \bibinfo {pages} {023819} (\bibinfo {year} {2015})}\BibitemShut
  {NoStop}%
\bibitem [{\citenamefont {Cabrera}\ \emph {et~al.}(2015)\citenamefont
  {Cabrera}, \citenamefont {Bondar}, \citenamefont {Jacobs},\ and\
  \citenamefont {Rabitz}}]{Cabrera15a}%
  \BibitemOpen
  \bibfield  {author} {\bibinfo {author} {\bibfnamefont {R.}~\bibnamefont
  {Cabrera}}, \bibinfo {author} {\bibfnamefont {D.~I.}\ \bibnamefont {Bondar}},
  \bibinfo {author} {\bibfnamefont {K.}~\bibnamefont {Jacobs}}, \ and\ \bibinfo
  {author} {\bibfnamefont {H.~A.}\ \bibnamefont {Rabitz}},\ }\href {\doibase
  10.1103/PhysRevA.92.042122} {\bibfield  {journal} {\bibinfo  {journal} {Phys.
  Rev. A}\ }\textbf {\bibinfo {volume} {92}},\ \bibinfo {pages} {042122}
  (\bibinfo {year} {2015})}\BibitemShut {NoStop}%
\bibitem [{\citenamefont {Kim}\ and\ \citenamefont {Kim}(2015)}]{Kim15a}%
  \BibitemOpen
  \bibfield  {author} {\bibinfo {author} {\bibfnamefont {K.-Y.}\ \bibnamefont
  {Kim}}\ and\ \bibinfo {author} {\bibfnamefont {S.}~\bibnamefont {Kim}},\
  }\href {\doibase 10.1016/j.sse.2015.04.007} {\bibfield  {journal} {\bibinfo
  {journal} {Solid State Electron.}\ }\textbf {\bibinfo {volume} {111}},\
  \bibinfo {pages} {22} (\bibinfo {year} {2015})}\BibitemShut {NoStop}%
\bibitem [{\citenamefont {Moyal}(1949)}]{Moyal49a}%
  \BibitemOpen
  \bibfield  {author} {\bibinfo {author} {\bibfnamefont {J.~E.}\ \bibnamefont
  {Moyal}},\ }\href {\doibase 10.1017/S0305004100000487} {\bibfield  {journal}
  {\bibinfo  {journal} {Math. Proc. Cambridge Phil. Soc.}\ }\textbf {\bibinfo
  {volume} {45}},\ \bibinfo {pages} {99} (\bibinfo {year} {1949})}\BibitemShut
  {NoStop}%
\bibitem [{\citenamefont {Rosati}\ and\ \citenamefont
  {Rossi}(2014)}]{Rosati14b}%
  \BibitemOpen
  \bibfield  {author} {\bibinfo {author} {\bibfnamefont {R.}~\bibnamefont
  {Rosati}}\ and\ \bibinfo {author} {\bibfnamefont {F.}~\bibnamefont {Rossi}},\
  }\href {\doibase 10.1103/PhysRevB.89.205415} {\bibfield  {journal} {\bibinfo
  {journal} {Phys. Rev. B}\ }\textbf {\bibinfo {volume} {89}},\ \bibinfo
  {pages} {205415} (\bibinfo {year} {2014})}\BibitemShut {NoStop}%
\bibitem [{Note1()}]{Note1}%
  \BibitemOpen
  \bibinfo {note} {The diagonal terms of the density matrix describe the
  population of the generic single-particle state~$\alpha $ while the
  off-diagonal terms describe the quantum-mechanical phase coherence (or
  polarization) between states $\alpha _1$ and $\alpha _2$.}\BibitemShut
  {Stop}%
\bibitem [{\citenamefont {Davies}(1976)}]{b-Davies76}%
  \BibitemOpen
  \bibfield  {author} {\bibinfo {author} {\bibfnamefont {E.}~\bibnamefont
  {Davies}},\ }\href@noop {} {\emph {\bibinfo {title} {Quantum theory of open
  systems}}}\ (\bibinfo  {publisher} {Academic Press},\ \bibinfo {year}
  {1976})\BibitemShut {NoStop}%
\bibitem [{\citenamefont {Breuer}\ and\ \citenamefont
  {Petruccione}(2007)}]{b-Breuer07}%
  \BibitemOpen
  \bibfield  {author} {\bibinfo {author} {\bibfnamefont {H.}~\bibnamefont
  {Breuer}}\ and\ \bibinfo {author} {\bibfnamefont {F.}~\bibnamefont
  {Petruccione}},\ }\href@noop {} {\emph {\bibinfo {title} {The Theory of Open
  Quantum Systems}}}\ (\bibinfo  {publisher} {OUP Oxford},\ \bibinfo {year}
  {2007})\BibitemShut {NoStop}%
\bibitem [{\citenamefont {Lindblad}(1976)}]{Lindblad76a}%
  \BibitemOpen
  \bibfield  {author} {\bibinfo {author} {\bibfnamefont {G.}~\bibnamefont
  {Lindblad}},\ }\href {\doibase 10.1007/BF01608499} {\bibfield  {journal}
  {\bibinfo  {journal} {Commun. Math. Phys.}\ }\textbf {\bibinfo {volume}
  {48}},\ \bibinfo {pages} {119} (\bibinfo {year} {1976})}\BibitemShut
  {NoStop}%
\bibitem [{Note2()}]{Note2}%
  \BibitemOpen
  \bibinfo {note} {In addition to the relaxation dynamics (\ref {RTADMEf}) of
  the level population $f_\alpha $, the density-matrix RTA model in (\ref
  {RTADME}) describes the scattering-induced decay of the interlevel
  polarization known as decoherence process.\cite {Rossi02b}}\BibitemShut
  {NoStop}%
\bibitem [{\citenamefont {Zhan}\ \emph {et~al.}(2016)\citenamefont {Zhan},
  \citenamefont {Colom\'es},\ and\ \citenamefont {Oriols}}]{Zhan16a}%
  \BibitemOpen
  \bibfield  {author} {\bibinfo {author} {\bibfnamefont {Z.}~\bibnamefont
  {Zhan}}, \bibinfo {author} {\bibfnamefont {E.}~\bibnamefont {Colom\'es}}, \
  and\ \bibinfo {author} {\bibfnamefont {X.}~\bibnamefont {Oriols}},\ }\href
  {\doibase 10.1007/s10825-016-0875-5} {\bibfield  {journal} {\bibinfo
  {journal} {J. Comput. Electron.}\ }\textbf {\bibinfo {volume} {15}},\
  \bibinfo {pages} {1206} (\bibinfo {year} {2016})}\BibitemShut {NoStop}%
\bibitem [{\citenamefont {Dolcini}\ \emph {et~al.}(2013)\citenamefont
  {Dolcini}, \citenamefont {Iotti},\ and\ \citenamefont {Rossi}}]{Dolcini13a}%
  \BibitemOpen
  \bibfield  {author} {\bibinfo {author} {\bibfnamefont {F.}~\bibnamefont
  {Dolcini}}, \bibinfo {author} {\bibfnamefont {R.~C.}\ \bibnamefont {Iotti}},
  \ and\ \bibinfo {author} {\bibfnamefont {F.}~\bibnamefont {Rossi}},\ }\href
  {\doibase 10.1103/PhysRevB.88.115421} {\bibfield  {journal} {\bibinfo
  {journal} {Phys. Rev. B}\ }\textbf {\bibinfo {volume} {88}},\ \bibinfo
  {pages} {115421} (\bibinfo {year} {2013})}\BibitemShut {NoStop}%
\bibitem [{Note3()}]{Note3}%
  \BibitemOpen
  \bibinfo {note} {In spite of the strong formal similarity with the
  conventional Boltzmann transport theory, we stress that the generalized
  Wigner-function scattering rates in (\ref {Pinout}) are not necessarily
  positive-definite.}\BibitemShut {Stop}%
\bibitem [{\citenamefont {Badziag}(1985)}]{Badziag85a}%
  \BibitemOpen
  \bibfield  {author} {\bibinfo {author} {\bibfnamefont {P.}~\bibnamefont
  {Badziag}},\ }\href {\doibase 10.1016/0378-4371(85)90046-9} {\bibfield
  {journal} {\bibinfo  {journal} {Physica A}\ }\textbf {\bibinfo {volume}
  {130}},\ \bibinfo {pages} {565} (\bibinfo {year} {1985})}\BibitemShut
  {NoStop}%
\bibitem [{\citenamefont {Serimaa}\ \emph {et~al.}(1986)\citenamefont
  {Serimaa}, \citenamefont {Javanainen},\ and\ \citenamefont
  {Varr\'o}}]{Serimaa86a}%
  \BibitemOpen
  \bibfield  {author} {\bibinfo {author} {\bibfnamefont {O.~T.}\ \bibnamefont
  {Serimaa}}, \bibinfo {author} {\bibfnamefont {J.}~\bibnamefont {Javanainen}},
  \ and\ \bibinfo {author} {\bibfnamefont {S.}~\bibnamefont {Varr\'o}},\ }\href
  {\doibase 10.1103/PhysRevA.33.2913} {\bibfield  {journal} {\bibinfo
  {journal} {Phys. Rev. A}\ }\textbf {\bibinfo {volume} {33}},\ \bibinfo
  {pages} {2913} (\bibinfo {year} {1986})}\BibitemShut {NoStop}%
\end{thebibliography}

%

\end{document}